%% file: main.tex
\documentclass[10pt,conference]{IEEEtran}
\usepackage{cite}
\usepackage{amsmath,amssymb,amsfonts}
\usepackage{algorithmic}
\usepackage{graphicx}
\usepackage{textcomp}
\usepackage[dvipsnames]{xcolor}

\usepackage[hyphens]{url}
\usepackage{xfrac}
\usepackage{xspace}
\usepackage{booktabs}
\usepackage{tabularx}
\usepackage{comment}
\usepackage{multirow}
\usepackage{makecell} 
\usepackage{enumitem}
\usepackage{hhline}
\usepackage{balance}
\usepackage[all]{nowidow}

\usepackage[caption=false]{subfig}

\usepackage{hyperref}

\newcolumntype{L}{>{\raggedright\arraybackslash}X}
\newcolumntype{C}{>{\centering\arraybackslash}X}
\newcolumntype{R}{>{\raggedleft\arraybackslash}X}
\newcommand{\NAME}{IVE\xspace}
\newcommand{\mathpolyring}{\mathcal{R}_Q}
\newcommand{\polyring}{$\mathpolyring$\xspace}
\newcommand{\mathDB}{\mathtt{DB}}
\newcommand{\DB}{$\mathDB$\xspace}

\newcommand{\evk}{$\mathtt{evk}$\xspace}
\newcommand{\evks}{$\mathtt{evk}$s\xspace}
\newcommand{\query}{$\mathtt{Query}$\xspace}
\newcommand{\ours}{\NAME}
\newcommand{\firstdim}{$\mathtt{RowSel}$\xspace}
\newcommand{\initdim}{\firstdim}
\newcommand{\restdim}{$\mathtt{ColTor}$\xspace}
\newcommand{\expandquery}{$\mathtt{ExpandQuery}$\xspace}
\newcommand{\mathsubs}{\mathtt{Subs}}
\newcommand{\subs}{$\mathsubs$\xspace}
\definecolor{revcolorb}{RGB}{2, 107, 213}

\newcommand{\ctbfv}{$\mathbf{ct}_\text{BFV}$\xspace}
\newcommand{\ctrgsw}{$\mathbf{ct}_\text{RGSW}$\xspace}

\newcommand{\mathctbfv}{\mathbf{ct}_\text{BFV}}
\newcommand{\mathctrgsw}{\mathbf{ct}_\text{RGSW}}
\newcommand*{\Scale}[2][4]{\scalebox{#1}{$#2$}}
\newcommand*\circled[1]{\raisebox{.5pt}{\textcircled{\raisebox{-.9pt} {#1}}}}

\title{\huge IVE: An Accelerator for Single-Server Private Information Retrieval Using Versatile Processing Elements}

\newcommand\hpcaauthors{Sangpyo Kim\IEEEauthorrefmark{2}, Hyesung Ji\IEEEauthorrefmark{3}, Jongmin Kim\IEEEauthorrefmark{3}, Wonseok Choi\IEEEauthorrefmark{3}, Jaiyoung Park\IEEEauthorrefmark{3}, and {Jung Ho} Ahn\IEEEauthorrefmark{3}}
\newcommand\hpcaaffiliation{\IEEEauthorrefmark{2}CryptoLab Inc., \IEEEauthorrefmark{3}Seoul National University}
\newcommand\hpcaemail{spkim@cryptolab.co.kr, \{kevin5188, jongmin.kim, wonseok.choi, jeff1273, gajh\}@snu.ac.kr}

\author{
\IEEEauthorblockN{\hpcaauthors{}}
      \IEEEauthorblockA{
        \hpcaaffiliation{} \\
        \hpcaemail{}
      }
\vspace{-0.2in}
}

\begin{document}
\maketitle

%


\begin{abstract}

Private information retrieval (PIR) is an essential cryptographic protocol for privacy-preserving applications, enabling a client to retrieve a record from a server's database without revealing which record was requested.
Single-server PIR based on homomorphic encryption has particularly gained immense attention for its ease of deployment and reduced trust assumptions.
However, single-server PIR remains impractical due to its high computational and memory bandwidth demands.
Specifically, reading the entirety of large databases from storage, such as SSDs, severely limits its performance.
To address this, we propose IVE, an accelerator for single-server PIR with a systematic extension that enables practical retrieval from large databases using DRAM.
Recent advances in DRAM capacity allow PIR for large databases to be served entirely from DRAM, removing its dependence on storage bandwidth.
Although the memory bandwidth bottleneck still remains, multi-client batching effectively amortizes database access costs across concurrent requests to improve throughput. 
However, client-specific data remains a bottleneck, whose bandwidth requirements ultimately limits performance.
IVE overcomes this by employing a large on-chip scratchpad with an operation scheduling algorithm that maximizes data reuse, further boosting throughput.
Additionally, we introduce sysNTTU, a versatile functional unit that enhances area efficiency without sacrificing performance.
We also propose a heterogeneous memory system architecture, which enables a linear scaling of database sizes without a throughput degradation.
Consequently, IVE achieves up to 1,275$\times$ higher throughput compared to prior PIR hardware solutions.

\end{abstract}
\input{intro}
\input{background}
\input{contribution-2}
\input{contribution-3}
\input{contribution-4}
\input{evaluation}
\input{related}
\input{conclusion}
\input{acknowledgment}

\balance
\bibliographystyle{IEEEtranS}
\bibliography{refs}

\end{document}

%% file: intro.tex
\section{Introduction}
\label{sec:intro}

Private information retrieval (PIR) allows users to query a remote database (\DB) without revealing their query, offering a cryptographic basis for privacy-preserving access to public data.
Amid the rapid expansion of cloud computing and growing emphasis on data governance, PIR is emerging as a critical building block for privacy-preserving applications, such as web search, location-based services, contact tracing, and AI inference~\cite{osdi-2021-fastpir,osdi-2016-unobservable,popets-2016-riffle,security-2011-pir-tor,nsdi-2016-popcorn,iacr-2013-efficient,DBSec-2010-generalizing,ccs-2016-reporting-ad,ccs-2011-practical,ASPLOS-2024-gpupir}.

Among various PIR protocols, those based on homomorphic encryption (HE) stand out due to their general applicability and low communication costs~\cite{ccs-2024-respire, osdi-2021-fastpir, ccs-2021-onionpir, sp-2022-spiral, security-2021-mulpir, sp-2018-sealpir, usenixsec-2023-simplepir}. 
HE~\cite{asia-2017-ckks,toct-2014-bgv,jc-2020-tfhe} is an encryption scheme that enables direct computation on encrypted data, allowing a server to process PIR queries without decryption. 
Unlike other PIR protocols~\cite{eurocrypt-2014-dpfpir, sp-2015-riposte,acm-1998-pir, ccsw-2014-raidpir} requiring additional assumptions, complex infrastructures with multiple servers, or both, HE-based PIR relies on strong cryptographic guarantees to ensure privacy with only a single server. 
Its strong security has led to its gradual adoption in practical applications~\cite{tip-toe, AppleLiveCallerID, insPIRe}, as exemplified by Apple’s use in private visual search.

This simplicity, however, comes at the cost of heavy server-side computation.
The high computational complexity of HE operations incurs long retrieval latencies, limiting its practical use.
For example, state-of-the-art PIR protocols~\cite{ccs-2024-kspir, usenixsec-2023-simplepir, sp-2022-spiral, security-2024-ypir, crypto-2024-hintless, iacr-2024-whispir} take 1.1--18.6 seconds for retrieving a 1B--32KB record from an 8GB \DB on a CPU-based system~\cite{ccs-2024-kspir}.

Numerous acceleration studies have been conducted for HE, leveraging CPUs/GPUs~\cite{access-2021-demystify,wahc-2021-hexl, iiswc-2020-ntt, tches-2021-100x, hpca-2023-tensorfhe, cal-2024-PBS, arxiv-2024-cheddar} or custom FPGAs/ASICs~\cite{micro-2021-f1,hpca-2023-fab,isca-2022-bts,micro-2022-ark,isca-2022-craterlake,isca-2023-sharp}. 
They focus on number-theoretic transforms (NTTs), with an emphasis on bootstrapping~\cite{eurocrypt-2018-heaanboot,asiacrypt-2016-PBS,jc-2021-bgv-boot}, which dominate the runtime in typical HE workloads.

However, PIR's memory-intensive nature hinders its acceleration: as concealing the target record requires scanning the entire \DB, for large {\DB}s exceeding DRAM capacity, the low bandwidth of secondary storage devices (e.g., SSDs) significantly degrades performance.
This limitation motivated INSPIRE~\cite{isca-2022-inspire} to adopt in-storage ASICs to accelerate HE-based PIR. Unfortunately, even with such efforts, PIR remains impractical, requiring 36 seconds to retrieve a 288B entry from a 288GB \DB for anonymous communication~\cite{osdi-2021-fastpir}.

To overcome this limitation, we propose \NAME, an accelerator for single-server HE-based PIR with a systematic extension  to support large {\DB}s efficiently.
Technology scaling now allows modern hardware systems to support terabyte-scale DRAM configurations, which open up new opportunities to accelerate the retrieval process by providing \DB data with notably higher DRAM bandwidth.
Our in-depth analysis shows that, although the memory bandwidth bottleneck for scanning \DB persists even with DRAM, \emph{batching multi-client queries} can amortize the overheads, improving PIR throughput. 

However, batching cannot resolve the bandwidth demands caused by client-specific data, limiting further improvements.
\NAME addresses this challenge by using a large on-chip scratchpad, combined with algorithmic optimizations that maximize on-chip data reuse during PIR. 
Moreover, leveraging the sequential computational patterns inherent in PIR, we develop a versatile functional unit, \emph{sysNTTU}, that consolidates multiple core PIR operations into a unified hardware block, boosting area efficiency without degrading PIR performance.

Further, we propose a scale-up systematic extension that leverages two heterogeneous memory types: higher-bandwidth HBM for data being processed and larger but lower-bandwidth LPDDR for offloading the full \DB.
We observe that, batching effectively amortizes \DB access costs to render potential PIR throughput degradations caused by low LPDDR bandwidth to be minimal.
For further {\DB} size scaling, we also introduce a scale-out system that connects multiple scale-up \NAME systems.
Our optimized parallelization strategy efficiently partitions \DB across these systems, enabling near-linear scaling of PIR throughput.
As a result, \NAME delivers up to 1,275$\times$ higher throughput over the state-of-the-art hardware solution for HE-based PIR~\cite{isca-2022-inspire} in practical applications~\cite{osdi-2021-fastpir,sp-2018-sealpir,popets-2016-xpir}.

In this paper, we make the following key contributions:
\begin{itemize}[leftmargin=*]
    \item We identify that batching multiple clients' requests exposes new opportunities for PIR acceleration on recent hardware by mitigating the bandwidth bottleneck caused by full-\DB scan. However, batching cannot solely address all the bandwidth demands due to client-specific data.
    \item We propose \NAME, a PIR accelerator with a large scratchpad and algorithmic techniques that maximize the on-chip reuse of client-specific data. \NAME effectively mitigates the memory bandwidth bottleneck to significantly enhance throughput. \NAME includes a versatile functional unit, \emph{sysNTTU}, that improves area efficiency without performance degradation.
    \item We offer a scalable, systematic solution that leverages a heterogeneous memory system, enabling \NAME to support PIR over large {\DB}s in practice.
\end{itemize}

%% file: background.tex
\section{Background}
\label{sec:background}

The practicality of various PIR schemes~\cite{usenixsec-2023-simplepir, ccs-2024-respire, osdi-2021-fastpir, sp-2022-spiral, ccs-2021-onionpir} is hindered by high computational and communication costs.
To reduce the costs, numerous studies employed additional assumptions.
Multi-server PIRs~\cite{eurocrypt-2014-dpfpir, sp-2015-riposte, acm-1998-pir, ccsw-2014-raidpir} assume that multiple non-colluding servers cooperate to process a client's query.
However, their underlying assumption is fragile as servers may collude through side channels and covert communication mechanisms~\cite{usenixsec-2004-tor, jcrypol-1988-dcnet}.
Also, establishing and maintaining a non-colluding-server environment is operationally complex.
These challenges make multi-server PIR less viable in practice~\cite{ccs-2024-respire, osdi-2021-fastpir, ccs-2021-onionpir, usenixsec-2023-simplepir, popets-2023-frodopir}.

Thus, \emph{we focus on simple (yet viable and useful) single-server PIR schemes.}
Table~\ref{tab:notation} summarizes the major symbols.
All vectors are column vectors.
We use $1\text{TB} \!=\! 2^{10}\text{GB} \!=\! 2^{20}\text{MB} \!=\! 2^{30}\text{KB} \!=\! 2^{40}\text{B}$.

\subsection{Na\"ive one-dimensional PIR}
\label{sec:background:1d-pir}

For an \emph{unencrypted database} (\DB) with $D$ records residing at a server, PIR allows a client to retrieve a record $\mathDB[i^*]$ without revealing the index $i^*$.
A client encrypts the index $i^*$ and sends it to the server.
The server must compute $\mathDB[i^*]$ with the unencrypted \DB and the encrypted $i^*$ (ciphertexts, \textbf{ct}s).

Homomorphic encryption (HE), a cryptographic primitive allowing oblivious computations on ciphertexts encrypted with small errors, is a perfect fit.
Numerous studies~\cite{ccs-2024-respire, sp-2022-spiral, ccs-2021-onionpir, ccs-2024-kspir,usenixsec-2023-simplepir} have proposed efficient single-server PIR schemes using HE, where they mostly share common structures.

We first describe the structure of a na\"ive PIR.
HE has a linear property; we can add two \textbf{ct}s ($\text{Enc}(\mathcal{X})$ and $\text{Enc}(\mathcal{Y})$) or multiply unencrypted data ($\mathcal{Z}$), denoted as \emph{plaintext}, to a \textbf{ct}. 
\begin{itemize}
    \item $\text{Enc}(\mathcal{X})+\text{Enc}(\mathcal{Y})\rightarrow\text{Enc}(\mathcal{X} + \mathcal{Y})$
    \item $\mathcal{Z}\cdot\text{Enc}(\mathcal{Y})\rightarrow\text{Enc}(\mathcal{Z}\cdot\mathcal{Y})$
\end{itemize}
The client uses a one-hot representation to send $D$ ciphertexts $\mathbf{ct}[0], \cdots, \mathbf{ct}[D-1]$ encrypting 0 (if $i\neq i^*$ for $\mathbf{ct}[i]$) or 1 (if $i=i^*$).
Then, the server uses the linearity of HE to compute
\begin{equation}
\label{eq:naive-pir}
\textstyle\sum_{i=0}^{D-1}\mathDB[i]\cdot\mathbf{ct}[i] \rightarrow\text{Enc}(\mathDB[i^*])
\end{equation}
and returns the ciphertext encrypting $\mathDB[i^*]$ to the client.

HE-based PIR schemes start from this na\"ive implementation accessing the entire \DB; in fact, even for advanced schemes, \emph{accessing the entire \DB is a critical requirement for PIR}~\cite{acm-1998-pir}.

\setlength{\tabcolsep}{3pt}
\begin{table}[t]
\centering
\caption{Symbols and their values we use for evaluation.}
\label{tab:notation}
\begin{tabularx}{0.999\columnwidth}{c|X|c}
\toprule
  Sym. & Meaning & Value\\
\midrule
  $D$ & Number (\#) of records in the database \DB. & $2^{16}$--$2^{24}$\\
  $D_0$ & Initial dimension size. & $2^8$\\
  $d$ & \# of subsequent dimensions. $D=D_0\cdot2^d$. & $8$--$16$\\
  $N$ & Degree of $\mathpolyring=\mathbb{Z}_Q[X]/(X^N + 1)$. & $2^{12}$\\
  $Q$ & Ciphertext modulus of \polyring. & $<2^{112}$\\
  $P$ & Plaintext modulus of $\mathcal{R}_P$. & $2^{32}$\\
  $z$ \& $\ell$ & Decomposition base and length. $z^\ell \ge Q$. & $2^{14}$--$2^{22}$ \& $5$--$8$\\
\bottomrule
\end{tabularx}
\end{table}
\setlength{\tabcolsep}{6pt}

\noindent \textbf{PIR based on Regev encryption:}\
Eq.~\ref{eq:naive-pir} can be implemented with Regev encryption~\cite{acm-2009-regev}.
A representative PIR scheme, SimplePIR~\cite{usenixsec-2023-simplepir}, is based on this method.
Each $\mathbf{ct}_\text{Regev}$ encrypts a single integer in $\mathbb{Z}_P$ for a plaintext modulus $P$ (i.e., an integer in a range of $[0, P-1]$), obtaining a length-$n$ (e.g., $n\!=\!2^{10}$ in SimplePIR) vector of integers in $\mathbb{Z}_Q$ for a ciphertext modulus $Q$ ($Q\gg P$) as the encryption result.
Here, $Q$ represents the budget for error amplification. 
Each HE operation amplifies the errors in a ciphertext, eventually making them undecryptable. 
Increasing $Q$ allows more computations on a ciphertext without requiring extra error mitigation procedures.
Each $\mathDB[i]$ is an unencrypted scalar in $\mathbb{Z}_P$.
After the client sends $D$ $\mathbf{ct}_\text{regev}$ ($nD\log Q$ bits of data) for the one-hot representation of $i^*$, the server computes Eq.~\ref{eq:naive-pir} by scalar-vector multiplications (mults) and vector-vector additions.

\begin{figure}
     \centering
     \includegraphics[width=0.84\columnwidth]{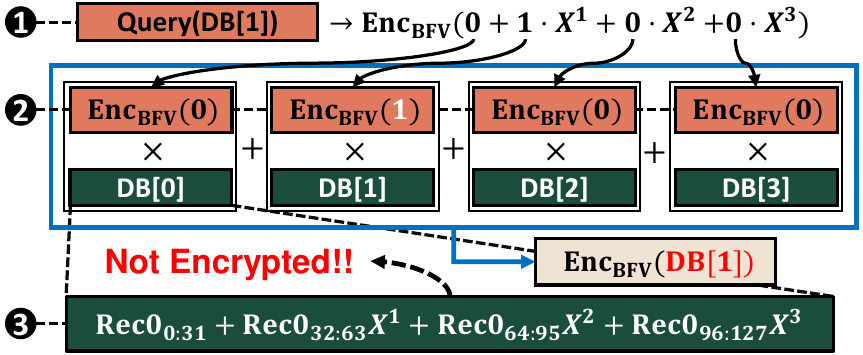}
     \caption{Na\"ive PIR based on BFV encryption~\cite{arxiv-2012-FV,crypto-2012-bfv} when $D \!=\! N\!=\!4$.}\label{fig:naive_PIR_protocol}
\end{figure}

\noindent \textbf{PIR based on BFV encryption:}\
To mitigate this massive data communication cost, numerous studies~\cite{ccs-2024-respire, ccs-2024-kspir, sp-2022-spiral, ccs-2021-onionpir} use BFV encryption~\cite{crypto-2012-bfv, arxiv-2012-FV} to represent $i^*$.
BFV can encrypt a degree-$(N\!-\!1)$ polynomial with its coefficients in a range of $[0, P - 1]$ ($\mathcal{R}_P=\mathbb{Z}_P[X]/(X^N+1)$).
Instead of sending $D$ \textbf{ct}s, the client prepares a single \emph{query} \textbf{ct} containing all bits for the one-hot index representation (see Fig.~\ref{fig:naive_PIR_protocol}-\circled{1}) by encrypting $X^{i^*}=0 + 0\cdot X + \cdots + 1 \cdot X^{i^*} + \cdots + 0 \cdot X^{N-1}$ (suppose $D = N$ for simplicity).
This process of combining multiple values in the range $[0, P - 1]$ into a polynomial is referred to as \emph{packing}.
The resulting BFV ciphertext, $\mathbf{ct}_\text{BFV}$, is a pair of polynomials in \polyring (e.g., $(a, b)\in\mathpolyring^2$) with a ciphertext modulus $Q$ (again, $Q\gg P$).
Thus, BFV substantially reduces the PIR communication cost ($2D\log Q$ bits of data) compared to that of Regev ($nD\log Q$ bits).

\begin{figure}
     \centering
    \includegraphics[width=\columnwidth]{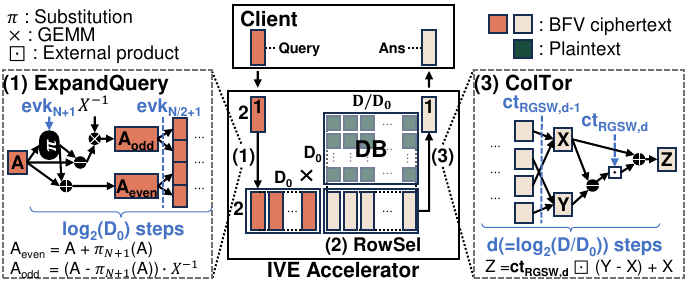}
     \caption{Server-side PIR computation process composed of (1) \expandquery, (2) \firstdim, and (3) \restdim.}\label{fig:PIR_protocol}
\end{figure}

\noindent \textbf{\expandquery, \subs, and \evks:}\
From a query encrypting $X^{i^*}$, the server can generate $D$ ciphertexts corresponding to the one-hot representation of $i^*$ (see Fig.~\ref{fig:naive_PIR_protocol}-\circled{2}) by performing a sequence of substitution (\subs) operations~\cite{sp-2018-sealpir}.
$\mathsubs(\mathbf{ct}_\text{BFV}, r)$ replaces $X$ with $X^r$ (e.g., $\text{Enc}_{\text{BFV}}(X+X^2) \rightarrow \text{Enc}_{\text{BFV}}(X^r+X^{2r})$) for the polynomial $\mathbf{ct}_\text{BFV}$ encrypts.
From $\mathtt{Query}=\text{Enc}_\text{BFV}(c_0+c_1X+\cdots+c_{N-1}X^{N-1})$, we can extract two BFV ciphertexts, each encrypting the odd/even terms.
We use $\mathsubs(\mathtt{Query},N+1)$ to negate the odd terms ($X^{2k+1}$) as $(X^{N+1})^{2k+1}=  -X^{2k+1}$ ($\because X^N=-1$ in \polyring):
\begin{equation*}
\Scale[0.85]{
\begin{aligned}
    &\mathsubs(\mathtt{Query}, N\!+\!1)\rightarrow\text{Enc}_\text{BFV}(c_0-c_1X+c_2X^2-\cdots-c_{N\!-\!1}X^{N\!-\!1})\\
    &\mathtt{Query}+\mathsubs(\mathtt{Query},N\!+\!1) \rightarrow \text{Enc}_\text{BFV}(c_0+c_2X^2+\cdots)\\
    &X^{-1}\cdot(\mathtt{Query}-\mathsubs(\mathtt{Query},N\!+\!1)) \rightarrow \text{Enc}_\text{BFV}(c_1+c_3X^2+\cdots)
\end{aligned}}
\end{equation*}
We elaborate on the computation of \subs in \S\ref{sec:background:rlwe-and-rgsw}.

The server iteratively uses the previous extraction results and performs \subs with $r=\sfrac{N}{2}+1,\sfrac{N}{4}+1,\cdots$ to obtain $\text{four}, \text{eight}, \cdots$ BFV ciphertexts containing $\sfrac{1}{4}, \sfrac{1}{8}, \cdots$ of the total terms in \query.
We refer to the whole process as \expandquery, which forms a binary-tree computational flow as shown in Fig.~\ref{fig:PIR_protocol}.
For each $r$ value in \subs, a unique public key $\mathtt{evk}_r$ needs to be prepared by the client.
Thus, \expandquery needs a distinct $\mathtt{evk}_r$ per tree depth, requiring up to $\log N$ \evks in total.
Eventually, the server obtains $\mathbf{ct}[i]=\text{Enc}_\text{BFV}(c_i)$ ($i=0,\cdots,D-1$), used to compute Eq.~\ref{eq:naive-pir}.

\subsection{Computational optimizations for Eq.~\ref{eq:naive-pir}}
\label{sec:background:computational_optimizations}

\noindent \textbf{Residue number system (RNS):}\
RNS simplifies handling $Q$ by decomposing it as $Q=q_0q_1q_2q_3$.
Chinese remainder theorem (CRT, Eq.~\ref{eq:crt}) is applied to replace each coefficient $c$ with four 28-bit residues ($[c]_{q_i}$) for moduli $q_i\in (2^{27}, 2^{28})$.
Inverse CRT (iCRT, Eq.~\ref{eq:icrt}) reconstructs $c$ from these residues.
\begin{align}
&\text{CRT: } ([c]_{q_0}, \cdots, [c]_{q_3})\gets (c\text{ mod }q_0, \cdots, c\text{ mod }q_3)\label{eq:crt}\\
&\text{iCRT: } c \gets \textstyle\sum_{i=0}^3 ([c]_{q_i} \cdot (\frac{Q}{q_i})^{-1} \text{ mod } q_i ) \cdot (\frac{Q}{q_i}) \text{ mod } Q\label{eq:icrt}
\end{align}
With RNS, a polynomial in \polyring becomes a length-$4N$ 28-bit vector (56KB when $N = 2^{12}$ as in Table~\ref{tab:notation}), also viewed as the concatenation of coefficients from four length-$N$ polynomials in $\mathcal{R}_{q_i}$'s.
A polynomial addition/mult over \polyring is equivalent to four independent additions/mults over $\mathcal{R}_{q_i}$'s.
With RNS, a \ctbfv becomes a $2\times4N$ structure (112KB).

\noindent \textbf{Number-theoretic transform (NTT):}\
NTT, a Fourier-transform variant, reduces the $\mathcal{O}(N^2)$ computational cost of a polynomial mult in \polyring, which is equivalent to a negacyclic convolution of the corresponding coefficient vectors.
With RNS, NTT is applied separately to each $\mathcal{R}_{q_i}$, making it four independent $N$-point fast Fourier transforms (FFTs) with $\mathcal{O}(N \log N)$ complexity.
NTT reduces a polynomial mult under RNS to an element-wise mult between two length-$4N$ vectors.
The $4N$ modular mults can be performed independently, known as \emph{coefficient-level parallelism} (CLP).
Polynomials are typically kept in their NTT form to support repeated mults without converting back by inverse NTT (iNTT).

\noindent \textbf{Preprocessing \DB:}\
BFV-based PIRs typically represent each $\mathDB[i]$ using a large plaintext domain ($\mathcal{R}_P$).
For a record $\mathbf{Rec0}$ with $N\log P = 32N$ bits of information (assuming $P=2^{32}$ from Table~\ref{tab:notation}), we can simply reinterpret $\mathbf{Rec0}$ as $N$ chunks of 32-bit data and set $\mathDB[0]=\mathbf{Rec0}_{0:31}+\mathbf{Rec0}_{32:63}X + \cdots + \mathbf{Rec0}_{32(N-1):32N-1}X^{N-1}$ (see Fig.~\ref{fig:naive_PIR_protocol}-\circled{3}) for  $\mathbf{Rec0}_{i:j}$ representing the data from the $i$-th bit to the $j$-th bit of $\mathbf{Rec0}$.
For smaller record sizes, we can construct $\mathDB[i]$ by packing multiple records into a single polynomial.

Then, calculating Eq.~\ref{eq:naive-pir} requires $D$ polynomial mults involving $\mathDB[i]$ for $D=\frac{\mathrm{DB_{size}}}{NlogP}$.
To facilitate this, we preprocess $\mathDB[i]$ to apply CRT and NTT in advance, after which $\mathDB[i]$ is in \polyring instead of $\mathcal{R}_P$.
This requires $\sfrac{\log Q}{\log P}\times$ ($<3.5\times$ based on Table~\ref{tab:notation}) more storage for the \DB.
Nevertheless, the preprocessing approach significantly speeds up PIR computation by more than 3.9$\times$ based on our CPU evaluation.

\subsection{Multi-dimensional PIR}
\label{sec:background:nd-pir}

The na\"ive implementation in \S\ref{sec:background:1d-pir} scales poorly with the \DB size ($D$) due to the $\mathcal{O}(D)$ communication cost.
Interpreting the \DB as a $(d + 1)$-dimensional structure ($D_0\times D_1\times\cdots\times D_d$) alleviates this problem.
Here, we use a three-dimensional example based on OnionPIR~\cite{ccs-2021-onionpir}, Spiral~\cite{sp-2022-spiral}, and Respire~\cite{ccs-2024-respire}.
Suppose the client wants to retrieve $\mathDB[i^*][j^*][k^*]$.

\noindent \textbf{Row selection (\initdim):}\
\expandquery can produce $\mathbf{ct}_\text{BFV}[0]$, $\mathbf{ct}_\text{BFV}[1]$, $\cdots$, $\mathbf{ct}_\text{BFV}[D_0 \!-\!1]$ corresponding to the one-hot representation of the initial dimension index $i^*$.
Then, we apply an Eq.~\ref{eq:naive-pir}-like method, denoted \firstdim, to compute
\begin{equation*}
\mathbf{ct}_{\text{BFV},\mathDB}^{(0)}[j][k] \gets \textstyle\sum_{i=0}^{D_0-1} \mathDB[i][j][k] \cdot \mathbf{ct}_\text{BFV}[i]\quad\forall j,k.
\end{equation*}
We obtain ``a row of \DB'' ciphertexts $\mathbf{ct}_{\text{BFV},\mathDB}^{(0)}$ with $D_1\cdot D_2$ entries each encrypting $\mathDB[i^*][j][k]$.

\noindent \textbf{RGSW encryption \& external product:}\
For $j^*$ and $k^*$, one might consider repeating the same process as \initdim.
However, with $\mathbf{ct}_{\text{BFV},\mathDB}^{(0)}$ given instead of unencrypted \DB, we need inter-ciphertext mults.
Early PIR work~\cite{security-2021-mulpir} directly performed inter-\ctbfv mults, introducing \emph{multiplicative increases in errors}.
It requires relatively larger parameters ($N$ and $Q$) to separate the error from data in a ciphertext.
This makes PIR computation and communication significantly more costly.

For a more lightweight PIR construction, OnionPIR~\cite{ccs-2021-onionpir} uses RGSW ciphertexts~\cite{crypto-2013-gsw} ({\ctrgsw}s) to perform \ctrgsw-\ctbfv mults, referred to as external products ($\boxdot$).
An external product produces a \ctbfv as an output ($\mathctrgsw\boxdot\mathctbfv\rightarrow \mathctbfv$) with \emph{just additive increases in error}, enabling the use of relatively small $N$ and $Q$ values to reduce the PIR cost.
Like \ctbfv, a \ctrgsw encrypts a polynomial in $\mathcal{R}_P$.
However, a \ctrgsw, which is composed of $4\ell$ polynomials ($2\times2\ell\times4N$ data structure, 1120KB for $\ell=5$), is $2\ell\times$ larger than a \ctbfv.
 
We can derive $D_1$ {\ctrgsw}s required for the one-hot representation of $j^*$ from a single \ctbfv by following the same process as \expandquery but with minor additional computations~\cite{devadas-2016-onionoram}.
Moreover, it is possible to pack all data required for $i^*, j^*, k^*$ into a single polynomial in $\mathcal{R}_P$, encrypt it into one \ctbfv, and perform \expandquery to extract the necessary {\ctbfv}s/{\ctrgsw}s for all dimensions at once.

\noindent \textbf{Practical implementation \& column tournament (\restdim):}\
With $D_1$ {\ctrgsw}s corresponding to the one-hot representation of $j^*$, we can use a \initdim-like accumulation method, albeit using external products instead.
Reducing the $D_1\!\cdot\!D_2$ entries in $\mathbf{ct}_{\text{BFV},\mathDB}^{(0)}$ to $D_2$ would require $D_1\!\cdot\!D_2$ external products for $j^*$.
However, the external product is much more computationally expensive than the polynomial-\ctbfv mults in \initdim (described in \S\ref{sec:background:rlwe-and-rgsw}).
Thus, it is beneficial to reduce the size of the subsequent dimensions.
In practice, $D_1=D_2=2$ and a high $D_0$ value would be used~\cite{sp-2022-spiral} for our three-dimensional example.
Further, when $D_1=2$, we need only one RGSW ciphertext $\mathbf{ct}_{\text{RGSW},j^*}$ directly encrypting $j^*$ (0 or 1)~\cite{ccs-2024-respire}.
To select \DB entries for $j^*$, we compute the following:
\begin{align*}
&\mathbf{ct}_{\text{BFV}, \text{diff}}[k]\gets\mathbf{ct}_{\text{BFV}, \mathDB}^{(0)}[1][k] - \mathbf{ct}_{\text{BFV}, \mathDB}^{(0)}[0][k],\\
&\mathbf{ct}_{\text{BFV}, \mathDB}^{(1)}[k] \gets \mathbf{ct}_{\text{RGSW},j^*} \boxdot \mathbf{ct}_{\text{BFV}, \text{diff}}[k] + \mathbf{ct}_{\text{BFV}, \mathDB}^{(1)}[0][k]\quad \forall k.
\end{align*}
Likewise, we finally compute the following for $k^*$:
\[
\mathbf{ct}_{\text{BFV}, \mathDB}^{(2)} \!\gets\! \mathbf{ct}_{\text{RGSW},k^*} \!\boxdot (\mathbf{ct}_{\text{BFV}, \mathDB}^{(1)}[1] - \mathbf{ct}_{\text{BFV}, \mathDB}^{(1)}[0]) + \mathbf{ct}_{\text{BFV}, \mathDB}^{(1)}[0].
\]

This process after \initdim is called \restdim.
As shown in Fig.~\ref{fig:PIR_protocol}, \restdim halves the number of selected \DB ciphertexts for every dimension.
The final result $\mathbf{ct}_{\text{BFV},\mathDB}^{(2)}$ encrypts $\mathDB[i^*][j^*][k^*]$.
\restdim also follows a binary-tree computational flow, with each tree depth corresponding to a dimension.

\noindent \textbf{Error analysis:}\
As an external product introduces additive error increases, the error~\cite{ccs-2021-onionpir} included in the PIR response ciphertext ($\mathbf{ct}_\text{resp.}=\mathbf{ct}_{\text{BFV}, \mathDB}^{(2)}$ in the above example) is bounded as $\mathtt{Err}(\mathbf{ct}_\text{resp.}) \leq \mathtt{Err}(\mathbf{ct}_{\text{BFV},\mathDB}^{(0)}) + \mathcal{O}(d)\cdot \mathtt{Err}(\mathctrgsw)$.
Typically, $\mathtt{Err}(\mathctrgsw)$ is much smaller than $\mathtt{Err}(\mathbf{ct}_{\text{BFV},\mathDB}^{(0)})$, which is largely determined by $D_0$ and $P$.
As $d$ grows logarithmically with the \DB size, the error remains stable under fixed $D_0$ and $P$ values, allowing PIR for practically large {\DB}s without extra error control measures such as bootstrapping.

\subsection{HE operations}
\label{sec:background:rlwe-and-rgsw}

\noindent \textbf{Linear BFV operations:}\
Linear BFV operations are performed in a polynomial-wise manner: 
i.e., for $\mathctbfv=(a, b)\in\mathpolyring^2$, $\mathctbfv'=(a', b')\in\mathpolyring^2$, and $p\in\mathpolyring$, $p\cdot\mathctbfv+\mathctbfv'=(p\cdot a+a', p \cdot b+b')$.

\begin{figure}[t]
    \centering
    \includegraphics[width=0.73\columnwidth]{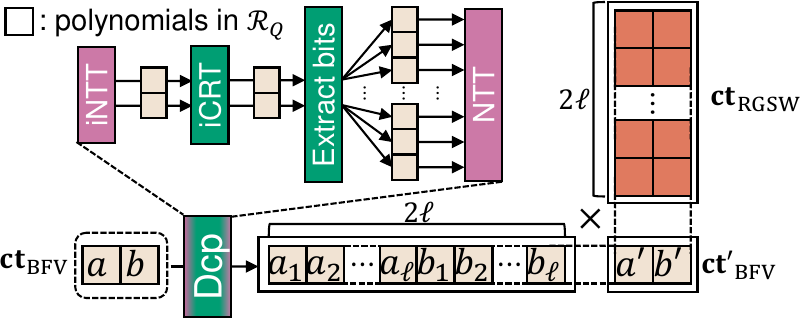}
    \caption{Computational flow of an external product ($\boxdot$).}
    \label{fig:external_product}
\end{figure}

\noindent \textbf{External product($\boxdot$):}\
For $\mathctrgsw  \boxdot \mathctbfv$, we first perform base decomposition (Dcp) to extend $a$ and $b$ of \ctbfv with a base $z$ (e.g., $2^{22}$). 
Dcp($a$) produces polynomials $a_{k\in[0,\ell)}$ (see Fig.~\ref{fig:external_product}), where each coefficient represents the $k$-th digit in base $z$ of the corresponding coefficient in $a$, falling within the range $[0, z-1]$. For Dcp($a$), we reconstruct each coefficient of $a$ with RNS by performing iNTT followed by iCRT (Eq.~\ref{eq:icrt}) and extract bits for $a_0,a_1,\cdots,a_{\ell-1}$. Similarly, we obtain $b_0,b_1,\cdots,b_{\ell-1}$. 
We reinterpret $a_k$ and $b_k$ as polynomials in $\mathpolyring$ under RNS with four moduli $q_{i\in[0,3]}$ and perform four NTT operations on each to facilite polynomial mults, resulting in a total of $4\times 2\ell$ NTTs. The decomposed polynomials of \ctbfv can then be multiplied with \ctrgsw as follows:
\[
\mathctrgsw \boxdot \mathctbfv = \mathctrgsw \cdot (\text{Dcp}(a), \text{Dcp}(b)) =\mathctbfv',
\]
\[
\text{Dcp}(x) = (x_0,x_1,\cdots,x_{\ell-1})\in \mathpolyring^\ell,\quad x = {\textstyle\sum_{i=0}^{\ell - 1}}\ x_i\cdot z^i.
\]
We regard a \ctrgsw as a $2\times2\ell$ matrix of polynomials and perform matrix-vector mult with the length-$2\ell$ vector of polynomials $(a_0,\cdots,a_{\ell-1}, b_0, \cdots, b_{\ell-1})$, resulting in a length-$2$ vector of polynomials, which is the output $\mathctbfv'$.

\noindent \textbf{\subs\& automorphism:}\
\subs also has a similar computation process.
The main computational cost of \subs for $\mathctbfv=(a,b)$ comes from computing $\text{Dcp}(a_{\text{Aut}, r})$ (length-$\ell$ vector of polynomials) and multiplying it with $\mathbf{evk}_r$ ($2\times\ell$ matrix of polynomials, 560KB for $\ell\!=\!5$).
$a_{\text{Aut}, r}$ and $b_{\text{Aut}, r}$ are the polynomials obtained by automorphism on $a$ and $b$, which shuffles the order of coefficients according to $r$.
\begin{equation*}
\mathsubs(\mathctbfv, r) = \mathbf{evk}_r \cdot \text{Dcp}(a_{\text{Aut}, r}) + (0, b_{\text{Aut}, r})
\end{equation*}

%% file: contribution-2.tex
\section{PIR Acceleration: Opportunities\&Challenges}
\label{sec:analysis}

Even with the high computational complexity of HE, accessing the full \DB renders PIR ill-suited for hardware acceleration.
For large SSD-resident DBs, storage latency dominates, motivating INSPIRE~\cite{isca-2022-inspire} to use in-storage ASICs with modest arithmetic throughput for PIR.

However, technology scaling brings new opportunities for PIR acceleration through increases in DRAM capacity.
Recent hardware systems now support TB-scale DRAM configurations~\cite{nvidiaB200Datasheet, supermicro2025nvl72}, enough to handle practical large {\DB} sizes.
This eliminates the need for low-bandwidth storage access during PIR, addressing the key bottleneck identified in INSPIRE.
Motivated by this, we explore an advanced hardware solution fully utilizing the high arithmetic throughput of modern hardware for HE-based PIR. 
We used an optimized variant of OnionPIR (see Fig.~\ref{fig:PIR_protocol}) as our main scheme.
OnionPIR shares structural similarities with numerous PIR schemes~\cite{sp-2022-spiral, ccs-2024-respire, sp-2018-sealpir, ccs-2024-kspir} and incorporates most of the core operations from other HE-based PIR approaches~\cite{ccs-2024-kspir, usenixsec-2023-simplepir, OnionPIRv2, insPIRe}, making our acceleration strategy easily extendable to them (\S\ref{sec:evaluation:other_schemes}).

\begin{figure}
\centering
    \subfloat{\label{fig:db_complexity}}
    \hspace{0.005\columnwidth}
    \subfloat{\centering\includegraphics[width=0.89\columnwidth]{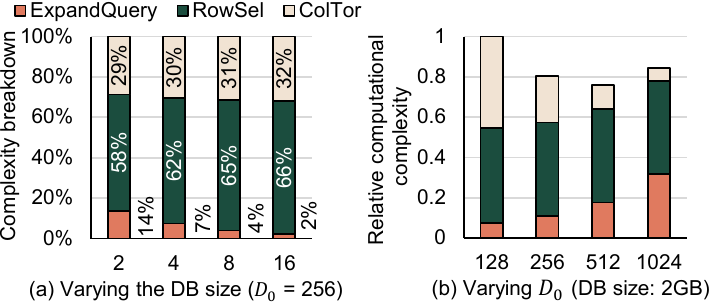}\label{fig:d_0_sweep}}
    \caption{Computational complexity breakdown based on the number of integer mults required for PIR depending on (a) the \DB size and (b) $D_0$.}\label{fig:breakdown}
\end{figure}

\subsection{Computational characteristics of PIR}
\label{sec:computational_PIR}

\begin{figure}
     \centering
     \includegraphics[width=0.9\columnwidth]{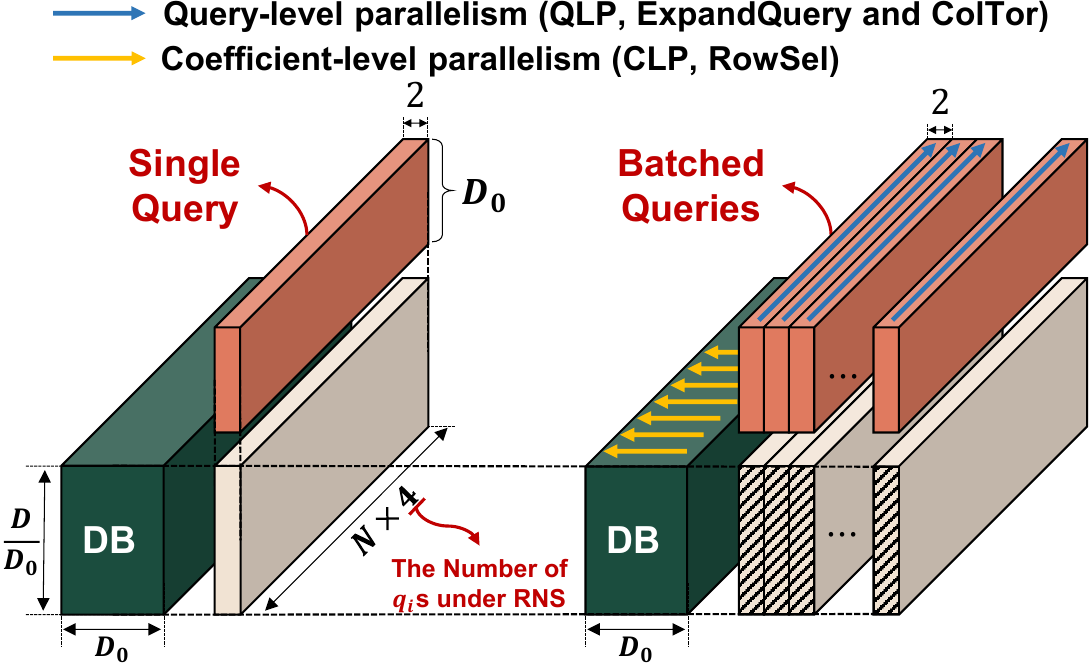}
     \caption{Computation of \firstdim without (left) and with (right) batching expressed as matrix-matrix mults.}
     \label{fig:batch}
\end{figure}

While \expandquery and \restdim both have a binary-tree computational flow with similar operations (\subs and external product), \firstdim only requires mults between \DB plaintexts and {\ctbfv}.
\firstdim can be expressed as a matrix-matrix mult (GEMM) between a $\frac{D}{D_0}\! \times\! D_0$ matrix (\DB) and a $D_0\! \times\! 2$ matrix ($D_0$ {\ctbfv}s), where each element is a polynomial in \polyring.
With RNS and NTT, a polynomial mult becomes an element-wise mult between two length-$4N$ vectors.
This allows us to reinterpret \firstdim as $4N$ independent GEMMs with modular integer arithmetic between a $\frac{D}{D_0} \times D_0$ matrix slice of the DB and a $D_0 \times 2$ slice of the query matrix (see Fig.~\ref{fig:batch}, left), where each element of a slice is a 28-bit integer.

Despite this simple structure, \firstdim is the most computationally demanding step.
A computational cost estimation based on the number of integer mults reveals that \firstdim accounts for 58--66\% of the total complexity for 2--16GB \DB (see Fig.~\ref{fig:db_complexity}).
Also, \firstdim dominates the total complexity for preferable $D_0$ values of 256--512, which minimize the total cost (see Fig.~\ref{fig:d_0_sweep}).

The large \DB size and poor \DB reusability become fundamental limiting factors as we attempt hardware acceleration for PIR.
As each \DB element is barely reused, limited DRAM bandwidth bottlenecks performance, hindering full use of abundant computing resources in modern hardware.

\subsection{Multi-client query batching}
\label{sec:batching}

\begin{figure}
\centering
    \includegraphics[width=\columnwidth]{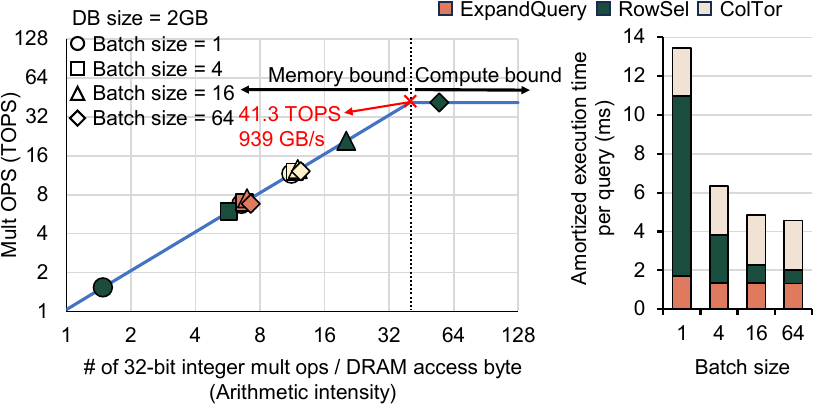}
    \caption{Theoretical roofline plot and execution time breakdown for each PIR step on RTX 4090 for various multi-client batch sizes (1--64) for a 2GB \DB.}\label{fig:roofline}
\end{figure}

We observe that batching PIR queries, possibly from multiple clients, presents a key opportunity for acceleration through hardware-friendly execution.
In general HE applications, batching is limited by client-specific cryptographic contexts that prevent data sharing.
By contrast, PIR is naturally amenable to batching as it operates over a shared, \emph{unencrypted} \DB.
Without batching, \firstdim is severely bottlenecked by memory bandwidth as it merely involves 1--2 integer mults per byte of DRAM access (Fig.\ref{fig:roofline}, left).
Batching mitigates this bottleneck by amortizing the \DB access cost across multiple queries, substantially increasing arithmetic intensity and enabling higher throughput.
Consequently, \firstdim throughput improves with larger batch sizes; thus, its share of the total PIR execution time decreases with larger batch sizes, as observed on an RTX 4090 GPU (Fig.~\ref{fig:roofline}, right).
Batching is applicable to most PIR schemes, not only HE-based ones, as they all require a full \DB scan that can be amortized through batching.

\begin{figure*}[tb!]
\centering
\subfloat{\label{fig:tree_a}}
\subfloat{\label{fig:tree_b}}
\subfloat{\label{fig:tree_c}}
\subfloat{\includegraphics[width=0.99\textwidth]{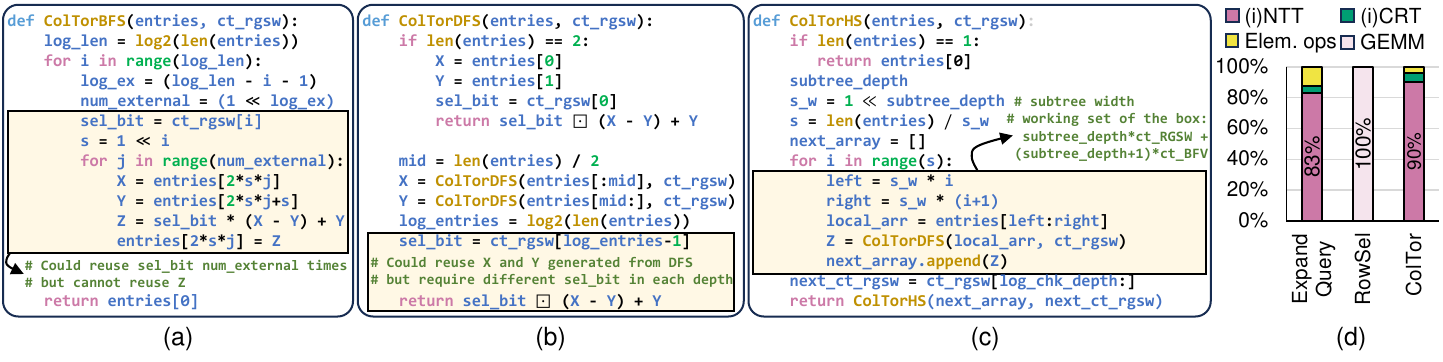}\label{fig:detail_complexity}}
\vspace{-0.1in}
  \caption{Operation scheduling of \expandquery, represented by a binary tree structure, when using 
   (a) breadth-first search (BFS), (b) depth-first search (DFS), and (c) hierarchical search, which can be used for both \expandquery and \restdim. (d) Computational complexity breakdown of three primary steps in PIR.}\label{fig:tree}
\end{figure*}

However, batching cannot alleviate the memory bandwidth bottleneck in the other steps. Since \expandquery and \restdim operate on client-specific data (e.g., \evk), the amortization cannot be applied, keeping the arithmetic intensity unchanged (see Fig.~\ref{fig:roofline}, left). 
As a result, these steps remain bounded by memory bandwidth due to their inherently low arithmetic intensity values.
Their execution time increases nearly linearly with the batch size, eventually accounting for non-negligible portions of the total PIR execution time (see Fig.~\ref{fig:roofline}, right).

%% file: contribution-3.tex
\section{Architecting an Accelerator for PIR}
\label{sec:architecture}

The memory bandwidth bottlenecks in \expandquery and \restdim limit the achievable PIR throughput of an accelerator as memory bandwidth scales more slowly than technology nodes.
Therefore, before detailing the architectural design of a PIR accelerator, we introduce several techniques designed to mitigate these bottlenecks within the accelerator.

\subsection{Optimizations for \expandquery and \restdim}
\label{subsec:optimizations}

To mitigate the DRAM access overhead during these steps, we introduce a hierarchical search (HS) algorithm to maximize data reuse during \expandquery and \restdim.
Common tree search algorithms, such as breadth-first search (BFS) and depth-first search (DFS) present distinct trade-offs for these steps. 
When using BFS for \restdim (see Fig.~\ref{fig:tree_a}), \ctrgsw (sel\_bit) can be highly reused across external products at the same depth, reducing \ctrgsw access overhead.
However, limited on-chip memory forces temporary outputs (Z) to be written back to and reloaded from DRAM for each subsequent depth, incurring significant memory access.
In contrast, DFS-based \restdim (see Fig.~\ref{fig:tree_b}) reduces DRAM access for temporary outputs by immediately processing intermediate results (X and Y) from prior depth. 
However, as a separate \ctrgsw is required for each depth, its reuse becomes severely limited.

HS combines the advantages of BFS and DFS while mitigating their drawbacks.
HS partitions the entire tree into subtrees whose working sets fit in the on-chip memory during processing, improving the reuse rate of \ctrgsw (or \evk) and \ctbfv.
For example, in \restdim using HS (see Fig.\ref{fig:tree_c}), the {\ctrgsw}s (ct\_rgsw) are reused across the entire loop and the temporary outputs (Z) are also reused in subsequent depth computations within each subtree.
This is enabled by selecting a proper subtree size that makes the working set fit in the on-chip memory.
Thus, HE requires only $2^{\mathrm{depth_{subtree}}}$ DRAM reads for the initial \ctbfv loads, and one DRAM store for the final result for each subtree.
This translates to a $\frac{2^{\mathrm{depth_{subtree}}}+1}{3\cdot2^{\mathrm{depth_{subtree}}}-3} \times$ reduction in DRAM access for \ctbfv over BFS.
Although HS reorders the operation scheduling sequences, it does not alter the order of operations applied to each ciphertext; thus, it does not introduce any additional error growth.

Our HS method favors a DFS-based approach for processing each subtree, rather than BFS, as DFS involves a smaller working set for a given subtree depth.
While larger subtree depth leads to greater reductions in DRAM access, increasing the depth also expands the working set, whose size can exceed the on-chip memory capacity.
In the \restdim utilizing BFS-based HS, the required memory capacity is given by
\[
\mathrm{depth_{subtree}}\cdot\mathctrgsw + 2^{\mathrm{depth_{subtree}-1}}\cdot\mathctbfv
\]
to accommodate all \ctbfv corresponding to the subtree width.
DFS-based approach, in contrast, computes external products as soon as possible, requiring a memory capacity of
\[
\mathrm{depth_{subtree}}\cdot\mathctrgsw + (\mathrm{depth_{subtree}}+1)\cdot\mathctbfv
\]
when one \ctbfv is waiting for tree reduction in each depth.
Thus, the DFS-based approach requires less space for \ctbfv;
conversely, it enables a larger subtree depth usage for a fixed on-chip memory capacity to maximize the benefits of HS.
HS is also applicable to \expandquery, as its computational process is a mirror image of \restdim.

To further mitigate the bottlenecks, we introduce \emph{reduction overlapping} (R.O.) for Dcp.
During an external product, Dcp expands a polynomial into $\ell$ polynomials, requiring a temporary memory space of $\ell \times$\ctbfv (see Fig.~\ref{fig:external_product}), which limits the HS subtree depth.
To address this, we overlap the reduction process with bit extraction and NTT computations for each $a_i$ and $b_i$.
By leveraging dedicated units for NTT and element-wise operations, intermediate polynomials are reduced just-in-time via partial GEMM, preventing them from occupying additional memory space.
R.O. is also applicable to \subs computations in \expandquery as they contain Dcp.

\begin{figure}
\centering
  \subfloat{\label{fig:expand_query}}
  \subfloat{\includegraphics[width=0.99\columnwidth]{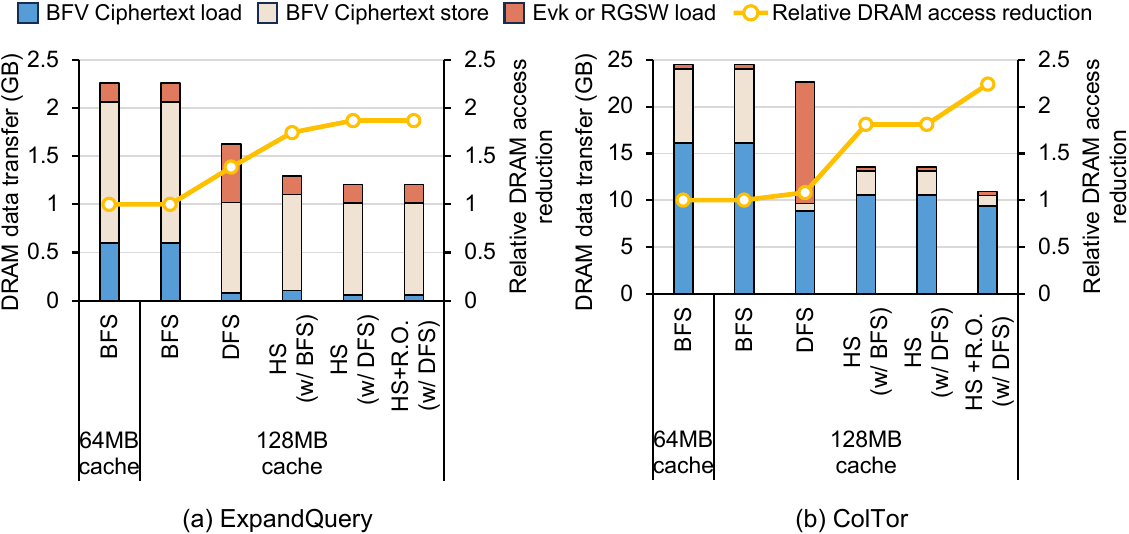}\label{fig:col_tor}}
  \vspace{-0.05in}
  \caption{Amounts of DRAM access during (a) \expandquery and (b) \restdim for 32-batched PIR queries for an 8GB \DB for different scheduling methods. R.O. denotes reduction overlapping.}\label{fig:data_reduction}
  \vspace{-0.05in}
\end{figure}

\begin{figure*}
     \centering
     \includegraphics[width=0.91\textwidth]{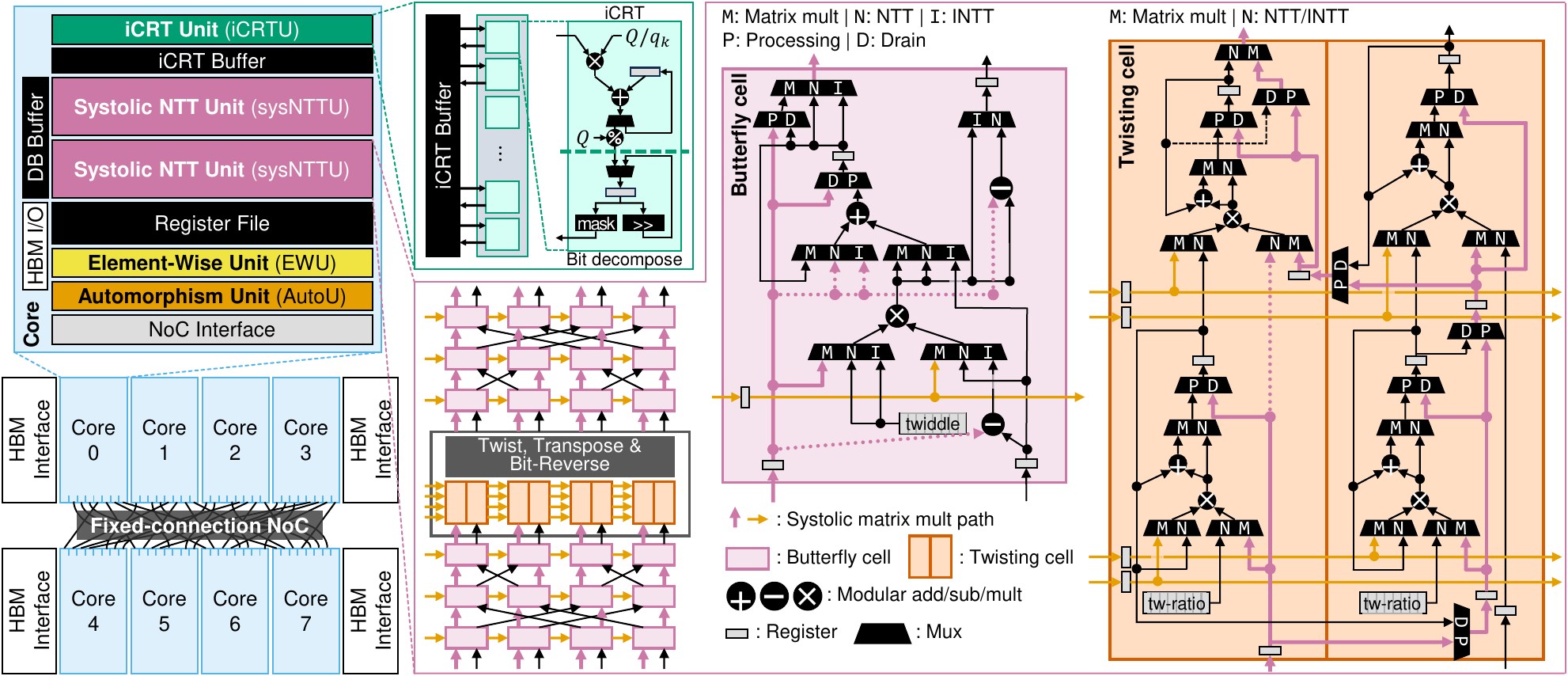}
     \caption{\ours architecture simplified to eight cores and eight lanes per core ($N=2^6$), and the microarchitectures of our systolic NTT unit (sysNTTU) and iCRT unit (iCRTU). The full \ours configuration features 32 cores and 64 lanes per core ($N=2^{12}$).}\label{fig:arch}
\end{figure*}

HS and R.O. significantly reduce DRAM access during \expandquery and \restdim, alleviating the bandwidth bottlenecks.
Compared to a BFS baseline, BFS-based HS achieves 1.75$\times$ and 1.81$\times$ reductions in DRAM data transfers for \expandquery and \restdim, respectively (Fig.~\ref{fig:data_reduction}).
DFS also reduces DRAM access over BFS, but its benefit is limited due to heavy access for \evks in \expandquery and \ctrgsw in \restdim.
DFS-based HS enables deeper subtree depths during \expandquery, yielding an extra 7\% reduction over BFS-based HS.
In \restdim, R.O. further shrinks working sets, supporting even larger subtree depths to reduce DRAM access by 1.23$\times$ over DFS-based HS.
Overall, our optimizations reduce DRAM traffic by 1.87$\times$ and 2.24$\times$ in \expandquery and \restdim, respectively (Fig.~\ref{fig:data_reduction}).
These benefits hold regardless of the \DB size as they depend only on the subtree configuration.
We further analyze their performance impacts in \S\ref{sec:sensitivity_study}.

\subsection{\NAME: An accelerator for batched single-server PIR}

Building on these optimizations, we propose \NAME, an efficient accelerator for batched single-server P\textbf{\underline{I}}R, leveraging \textbf{\underline{V}}ersatile processing \textbf{\underline{E}}lements.
\firstdim is dominated by GEMM, while other steps are NTT-heavy (Fig.\ref{fig:detail_complexity}), requiring support for both computations (\S\ref{sec:computational_PIR}).
However, as PIR steps execute sequentially (\S\ref{sec:arch:sysnttu}), hardware specialized for a single type of computation would often remain idle.
To address this inefficiency, \NAME introduces systolic NTT unit (sysNTTU), a versatile engine capable of performing both NTT and GEMM.
\ours comprises 32 vector cores interconnected via a network-on-chip (NoC) that adapts data movement to the parallelism of each step.
Each core follows a 64-lane vector core design from F1\cite{micro-2021-f1}, also adopted in \cite{isca-2022-craterlake, micro-2022-ark, isca-2023-sharp}.
Each core contains two sysNTTUs, an iCRT unit (iCRTU), an element-wise unit (EWU), an automorphism unit (AutoU), and 5MB of managed SRAM partitioned into register file (RF) and buffers.

\subsection{Versatile systolic NTTU (sysNTTU)}
\label{sec:arch:sysnttu}

Due to the strong data dependency between the steps, the PIR computation is inherently sequential (\expandquery $\rightarrow$ \firstdim $\rightarrow$ \restdim).
Pipelining offers no benefit as it increases data movement.
For example, consider pipelining \firstdim with \expandquery by chunking the $D_0$ BFV ciphertexts ($\mathctbfv[0],\cdots, \mathctbfv[D_0-1]$), which function as the output of \expandquery and the input for \firstdim, to start partial \firstdim computation right after \expandquery computation is finished for each chunk.
However, this approach incurs $(\text{\# of chunks})\times$ more memory access to accumulate partial outputs from \firstdim, which degrades PIR throughput.

We design our sysNTTU to support both (i)NTT and GEMM within a unified unit, instead of using separate functional units for them.
With the sequential processing of PIR, the sysNTTU improves hardware area efficiency without sacrificing performance.
Fig.~\ref{fig:arch} shows the microarchitecture of sysNTTU. 
We start from a fully pipelined vector NTT unit design from prior studies~\cite{micro-2021-f1, micro-2022-ark}, placing $\frac{\sqrt{N}}{2}\log N$ butterfly cells, which directly follow the FFT dataflow, with additional twisting, transposition, and bit-reversal circuits.
While mostly reusing the existing datapath, we add GEMM datapath based on the output-stationary systolic array dataflow~\cite{isca-2016-eyeriss} (see Fig.~\ref{fig:arch}).
Each butterfly cell also function as a systolic array cell.
Furthermore, a pair of twisting cells, which utilizes on-the-fly twisting~\cite{micro-2022-ark}, is repurposed as four systolic array cells.
The resulting sysNTTU function as a $\frac{\sqrt{N}}{2}\times(\log N + 4) = 32\times16$ systolic array, switching between NTT and GEMM via configurable MUXes.
For GEMM, the \DB buffer streams the database matrix horizontally, while the RF supplies the query matrix vertically (pink path in Fig.~\ref{fig:arch}).
Final outputs are drained through the query path by asserting the drain flag (D) on the MUXes.
Two sysNTTUs per core, operating at 1GHz, deliver 1TOPS of modular multiply-and-add (MMAD) throughput for GEMM, while also supporting (i)NTT with high throughput.

\begin{figure*}
\centering
  \centering
  \includegraphics[width=0.88\textwidth]{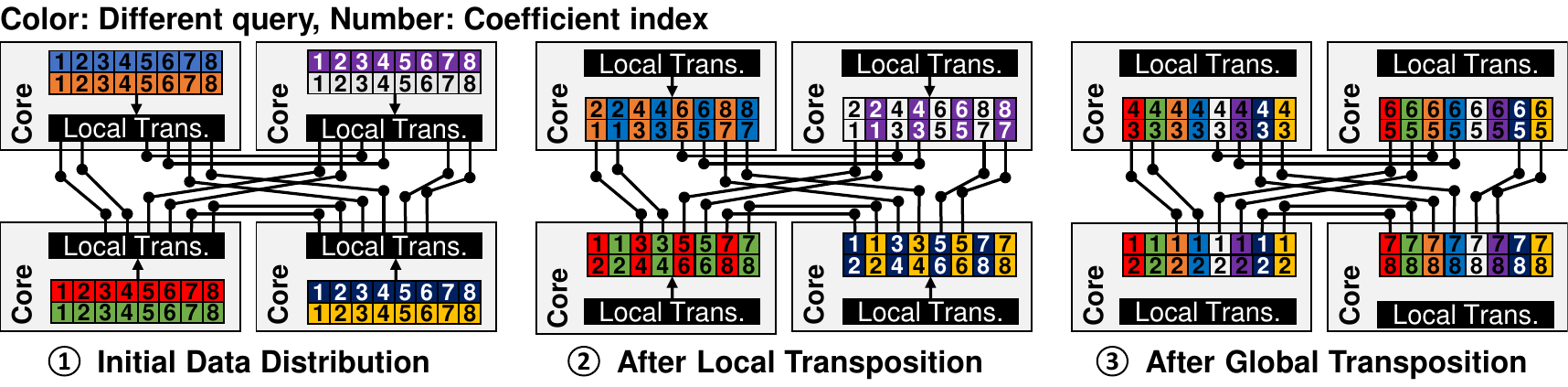}
  \vspace{-0.05in}
  \caption{Data exchange methods using the hierarchical network-on-chip (NoC) structure of \NAME simplified to four cores and eight lanes per core. Different colors represent different queries, while different numbers indicate distinct coefficients.}\label{fig:noc}
\end{figure*}

\subsection{Parallelization strategies}
\label{sec:analysis:parallelization}

Batched PIR execution exposes abundant parallelism that can be exploited by the cores in \NAME.
First, \firstdim can leverage CLP (\S\ref{sec:background:computational_optimizations}) as it can be regarded as a $4N$-parallel GEMMs (see Fig.~\ref{fig:batch});
NTT has transformed a polynomial mult into an element-wise mult between length-$4N$ coefficient vectors.
To exploit CLP, each set of $4N$ coefficients is evenly distributed across the cores of \NAME.
Then, each core computes $\frac{4N}{\text{\# of cores}}$ GEMM operations in parallel.

\expandquery and \restdim cannot exploit CLP as their (i)NTT and iCRT operations involve inter-coefficient computations.
Instead, we utilize another type of parallelism stemming from batching: \emph{query-level parallelism (QLP)}.
To exploit QLP, each core independently performs \expandquery and \restdim for different queries, preventing inter-core communication.

\subsection{Network-on-chip (NoC)}
\label{sec:arch:noc}

Due to the intersecting data distribution methods in the two parallelization strategies (\S\ref{sec:analysis:parallelization}), we need to `transpose' the data layout by swapping the `axis' for batches with that of coefficients.
To enable data layout transposition between adjacent PIR steps, \NAME leverages a hierarchical NoC structure with local data transpose units and fixed-wire global interconnects.
Fig.~\ref{fig:noc} exemplifies the transition from QLP to CLP.
At start, each core holds data from different queries for QLP (Fig.~\ref{fig:noc}-\circled{1}).
For the transition to CLP, \NAME first performs a local transposition at each core, loading a data element from each lane and transposing each $\frac{(\text{\# of lanes})}{(\text{\# of cores})} \times \frac{(\text{\# of lanes})}{(\text{\# of cores})}$ data block (Fig.~\ref{fig:noc}-\circled{2}).
The local transpose unit design is inspired by CraterLake~\cite{isca-2022-craterlake}.
Then, the cores can exchange data globally in a fixed transposition pattern, requiring each lane to connect to only one other lane in a different core through a fixed-wire global interconnect.
Overhead due to the global interconnects linearly increases with the number of cores, allowing us to scale the system for many cores with small NoC overheads.
After the global transposition, each core contains coefficients for the same coefficient index, but from (\# of lanes) different queries (Fig.~\ref{fig:noc}-\circled{3}), the data distribution of which fits CLP.

\subsection{Other hardware components}
\label{sec:other_hardware}

\noindent\textbf{iCRT unit (iCRTU):}\
Fig.~\ref{fig:arch} shows the microarchitecture of an iCRTU used in \NAME.
An iCRTU consists of $\sqrt{N}$ iCRTU cells, each responsible for two key computations required during Dcp: iCRT and bit extraction.
Each iCRTU cell operates independently, processing separate coefficients of the same polynomial in parallel.
To calculate Eq.~\ref{eq:icrt}, sysNTTUs are in charge of calculating $[a]_{q_k} \cdot (Q/q_k)^{-1} \text{ mod } q_k$ and each iCRTU cell is in charge of multiplying $(Q/q_k) \text{ mod } Q$.
This process reconstructs a coefficient of the original polynomial from its RNS representation.
The following bit extraction stage iteratively extracts least significant bits from the reconstructed coefficient through masking and right-shifting.

\noindent\textbf{Element-wise operation unit (EWU):}\
For each cycle, an EWU can handle basic element-wise operations, such as $\sqrt{N}=64$ MMAD operations, and also support small GEMMs between $2\times2$ and $2\times\sqrt{N}$ integer matrices used for external products and \subs.
We enable data forwarding from the sysNTTUs to the EWU; thus, the NTT results can be directly used for R.O., reducing RF bandwidth pressure.

\noindent\textbf{Automorphism unit (AutoU):}\
For automorphism, we adopted the fully pipelined design of ARK~\cite{micro-2022-ark}.

\noindent\textbf{Memory hierarchy:}\
IVE deploys three on-chip SRAM types: the main RF, an iCRT buffer, and a \DB buffer.
The RF, the largest at 4MB per core, stores ciphertexts and \evks for reuse.
Exploiting sequential $N$-granular access pattern, it employs wide ports (2 words per lane) and interleaved banks to maximize bandwidth. 
The 448KB iCRT and \DB buffers hold intermediate iNTT/iCRT results and \DB plaintexts for \initdim, respectively.
For an even bandwidth distribution and simplified access, each HBM channel is statically mapped to a core, leveraging the uniform data distribution across the cores.

\subsection{Exploiting special primes}
\label{sec:special_prime}

Due to the small number of primes needed for PIR, we leverage special primes resembling Solinas primes~\cite{techreport-1999-solinas} to reduce modular reduction circuit costs. 
It reduces area and power overhead by replacing costly mults with bit shifts. 
The scarcity of such primes prevented prior HE accelerators from adopting this strategy.
We use four primes with the form of $2^{27} + 2^k + 1$ ($k\in\{15,17,21,22\}$) for \ours.
Utilizing special primes reduces the area of a modular mult circuit based on Montgomery reduction by 9.1\% compared to the circuit based on primes satisfying $q_k=1\text{ mod } 2^{14}$~\cite{dsd-2019-word-level-montgomery} used in F1~\cite{micro-2021-f1}.

%% file: contribution-4.tex
\section{Designing Scalable Deployment Systems}
\label{sec:deployment_sys}

We enable scalability for large \DB sizes through a deployment system combining scale-up and scale-out approaches.

\begin{figure}
\centering
  \includegraphics[width=0.73\columnwidth]{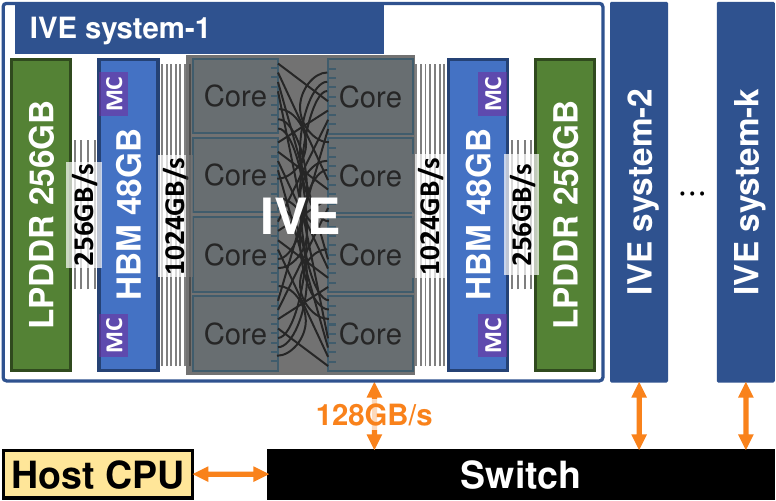}
  \caption{Architecture of the deployment system using scale-up and scale-out system of  \NAME.}\label{fig:depolyment_system}
\end{figure}

\noindent\textbf{Scale-up:}\
We propose a scale-up \emph{\NAME system}, which uses LPDDR memory as an expander to store \DB (see Fig.~\ref{fig:depolyment_system}).
While LPDDR provides lower bandwidth than HBM, it offers larger capacity\cite{HPCA-2024-CXLPNM}.
We integrate an LPDDR memory controller~\cite{lpddr_memeory_controller_patent} onto the logic die of custom HBM.
Custom HBM~\cite{ectc-2024-custom-hbm} leverages the logic die area for computation, enhancing chip area efficiency.
This approach is actively adopted in various research fields~\cite{hpca-2025-anaheim, micro-2024-duplex}.
Our design employs four 3D-stacked LPDDR modules~\cite{HPCA-2024-CXLPNM}, each providing 128GB of capacity and 128GB/s of bandwidth.

Our system adaptively uses HBM and LPDDR according to \DB size.
For \DB fitting in HBM, only HBM is used to avoid the latency overhead of accessing \DB from LPDDR.
A larger \DB exceeding the HBM capacity is offloaded to LPDDR and accessed in a streaming manner during \firstdim, while HBM continues to serve memory-bound \expandquery and \restdim steps.
Since \DB data is accessed only in \firstdim, where multi-query batching mitigates the bandwidth bottlenecks, the lower bandwidth of LPDDR has negligible impact on PIR throughput as batch size grows (later discussed with Fig.~\ref{fig:sensitivity_d}).
In our configuration, an \NAME system supports up to 128GB of \DB.

\noindent\textbf{Scale-out:}\
A scale-out \NAME system, denoted as \emph{\NAME cluster}, enables a linear scaling of supported \DB size by increasing the number of \NAME systems.
In an \NAME cluster, $\mathrm{num_{system}}$ \NAME systems are connected via a central PCIe switch, providing up to 128GB/s of interconnect bandwidth.
We introduce \emph{record-level parallelism (RLP)}, which enables parallel processing across the \NAME systems.
In RLP, the \DB matrix is partitioned along the $\frac{D}{D_0}$ dimension (see Fig.\ref{fig:batch}), with each node receiving a $\frac{D}{D_0\cdot \mathrm{num_{system}}} \times D_0 \times 4N$ matrix slice.
Each system performs \firstdim on its assigned \DB slice and can independently execute \restdim on the \firstdim output because the early stages of \restdim involve tournament-style selection among adjacent column entries within the local slice.
Finally, the partial \restdim outputs from systems are gathered into a single \NAME system for the final \restdim processing.
The gathering overhead is negligible (Comm. in Fig.~\ref{fig:sensitivity_d}) as each node sends only a single ciphertext after local \restdim.

\noindent\textbf{Batch scheduler:}\
To use multi-client batching in reality, we need a batch scheduler.
We employ a basic batch scheduler that has a time window waiting for queries, termed a \emph{waiting window}, to bound the latency overhead of batching.
We set the duration of this window based on the \DB access time in \initdim.
As batching only amortizes the cost of \DB access, waiting beyond this threshold increases latency without improving throughput.
In this way, the latency overhead remains below 2$\times$ while fully leveraging the benefits of multi-client query batching.
A detailed analysis of the latency overhead introduced by our batch scheduler is provided in \S\ref{sec:eval:batch_schedule}.

%% file: evaluation.tex
\section{Evaluation}
\label{sec:evaluation}

\subsection{Implementation and experimental setup}
\noindent\textbf{Hardware modeling:}\
We modeled IVE’s hardware cost primarily through RTL synthesis using the ASAP7 7.5-track 7nm predictive process design kit~\cite{mj-2016-asap7}.
SRAM and long wiring were evaluated with a modified version of FinCACTI~\cite{isvlsi-2014-fincacti}, incorporating published 7nm process data~\cite{isscc-2017-7nm-sram, iedm-2017-gf7nm, vlsit-2018-samsung, isscc-2018-7nm-sram-euv, iedm-2016-tsmc7nm, whitepaper-2018-irds, isca-2021-tpuv4i} for calibration.
The hardware components run at 1GHz (RF at 1.25GHz).
The RF, \DB buffer, and iCRT buffer use single-ported SRAM banks, providing 2.04TB/s, 0.81TB/s, and 0.41TB/s of per-core bandwidth, respectively.
Each 24GB HBM stack offers 512GB/s of bandwidth~\cite{isscc-2022-hynix-hbm3}, with four stacks used chip-wide.
DRAM power and HBM PHY area follow prior estimates~\cite{micro-2017-finegrainedDRAM, isca-2021-tpuv4i}. 
The default 32-core IVE configuration with 5MB SRAM per core occupies 155.3mm\textsuperscript{2} of area and consumes up to 239.1W of power (see Table~\ref{tab:asic-performance}), including HBM.

\noindent\textbf{Performance modeling:}\
We developed a cycle-level simulator that models each functional unit.
Given parameters, such as $D_0$, $d$, and batch size, as input, it constructs an operation graph for respecting data dependencies and avoiding hazards.
Operations are issued once dependencies are cleared, decomposed into core functions (e.g., (i)NTT), and dispatched to appropriate units.
Each functional unit maintains a separate queue and executes instructions based on pipeline availability.

\noindent\textbf{Data scheduling:}\
We adopted the decoupled data orchestration approach from CraterLake~\cite{isca-2022-craterlake}. 
As HE workloads form static directed acyclic computational graphs~\cite{micro-2021-f1}, the compiler can precompute an optimal schedule, prefetching data independently of the compute flow to hide memory latency.

\noindent\textbf{Compared systems:}\
We used the latest open-source implementations of OnionPIRv2~\cite{OnionPIRv2}, which is an optimized version of OnionPIR.
Performance was measured on an Intel Xeon Max 9460 CPU system with 1TB of DDR5-4800 memory.
Additionally, we implemented an OnionPIR-based PIR protocol on GPUs for computational analysis (\S\ref{sec:analysis}) and performance comparisons, using NVIDIA RTX4090 and H100 GPUs.
Our GPU implementation includes optimized kernel implementations of primitive functions and leverages both CLP and QLP across all kernels to maximize parallelism.

\noindent\textbf{PIR configuration:}\
Table~\ref{tab:notation} lists each parameter value used for evaluation.
This parameter setting is aligned with OnionPIR but adjusted for 128-bit security constraints~\cite{iacr-2024-guideline} by reducing $Q$.
For \NAME, a batch size of 64 was used unless otherwise specified.
For GPUs, the maximum batch size from the device memory capacity was used.
We used queries processed per second (QPS) as the primary metric for comparison.
We used both synthesized small {\DB}s and real workloads, including voice calling ($\mathtt{Vcall}$)~\cite{osdi-2021-fastpir}, anonymous communication ($\mathtt{Comm}$)~\cite{sp-2018-sealpir, osdi-2016-unobservable}, and file system ($\mathtt{Fsys}$)~\cite{popets-2016-xpir}.

\setlength{\tabcolsep}{6pt}
\begin{table}[t]
\centering
\caption{Area and peak power consumption of 32-core \NAME}\label{tab:asic-performance}
    \begin{tabularx}{0.8\columnwidth}{Lrr}
    \toprule
    \textbf{Component} & \textbf{Area (mm\textsuperscript{2})} & \textbf{Peak power (W)}\\
    \midrule
    sysNTTU & 0.77 & 2.17  \\
    iCRTU & 0.05 & 0.13  \\
    EWU  & 0.10 & 0.37  \\
    AutoU & 0.07 & 0.11  \\
    RF \& buffers  & 1.38 & 1.63 \\ 
    \midrule
    \textbf{1 core} & \textbf{2.91} & \textbf{5.12} \\
    \midrule
    32 cores & 93.1 & 163.8 \\
    NoC & 2.6 & 6.7 \\
    HBM & 59.6 & 68.6 \\
    \midrule
    \textbf{Sum} & \textbf{155.3} & \textbf{239.1} \\
    \bottomrule
    \end{tabularx}
\end{table}
\setlength{\tabcolsep}{6pt}

\subsection{Performance evaluation}

\begin{figure*}[t]
\centering
    \subfloat{\label{fig:sensitivity_a}}
    \subfloat{\label{fig:sensitivity_b}}
    \subfloat{\label{fig:sensitivity_c}}
    \subfloat{\label{fig:sensitivity_d}}
    \subfloat{\includegraphics[width=0.98\textwidth]{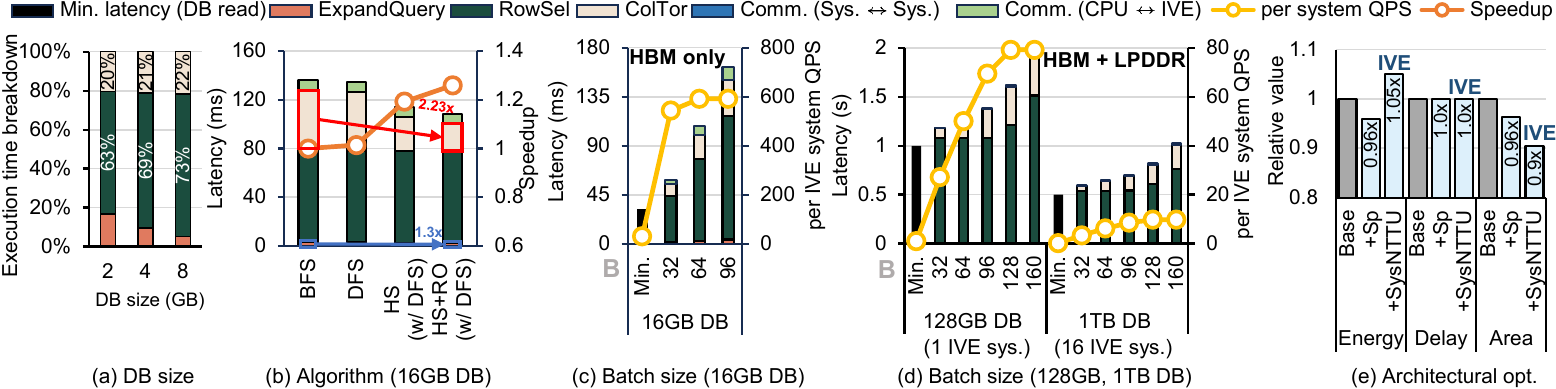}\label{fig:sensitivity_e}}
  \caption{Sensitivity study of \NAME with (a) different database sizes, (b) different algorithms, (c) batch size scaling for 16GB \DB, (d) batch sizes scaling for large-scale {\DB}s (128GB and 1TB), and (e) architectural optimizations.}\label{fig:sensitivity}
\end{figure*}

\NAME achieves superior throughput than both prior studies and our GPU implementation across all evaluated \DB sizes.
In terms of geometric mean (gmean), \NAME achieves $687.6\times$ higher QPS than 32-core CPU implementations for 2--8GB {\DB}s.
GPUs outperform CPUs due to higher DRAM bandwidth and integer computation throughput, even without batching.
While batching further improves QPS on GPUs, \NAME achieves up to $18.7\times$ higher average throughput compared to the best GPU-based batched PIR.
Finally, as \NAME mitigates the bandwidth bottlenecks in \expandquery and \restdim, each step's execution time (Fig.~\ref{fig:sensitivity_a}) closely aligns with its computational complexity (Fig.~\ref{fig:db_complexity}), exhibiting compute-bound characteristics.

Further, \NAME consume less energy than the baseline platforms.
We measured energy of the CPU and the GPU using Intel’s Running Average Power Limit (RAPL)~\cite{intel_rapl} and NVIDIA Management Library (NVML) API~\cite{nvidia-nvml}, respectively.
We estimated \NAME's total energy consumption based on each component's utilization.
CPUs consumed 72J, 107J, and 176J per query for 2GB, 4GB, and 8GB {\DB}s, respectively.
The GPU implementation with batching showed, on average, 43.3$\times$ lower energy consumption compared to the CPU across all database sizes.
Finally, \NAME, achieving the shortest latency, consumed 0.03J, 0.05J, and 0.09J per query for 2GB, 4GB, and 8GB databases, respectively, resulting in an average of 51.3$\times$ energy consumption reduction compared to the GPU.

IVE outperforms state-of-the-art PIR hardware acceleration studies as shown in Table~\ref{tab:comparison}.
Compared to CIP-PIR~\cite{usenixsec-2022-gpupir} and DPF-PIR~\cite{github-gpudpf}, both of which utilize GPUs to accelerate multi-server PIR schemes, \NAME processes 74.1$\times$ and 5.0$\times$ higher throughput in gmean, respectively, while providing stronger security guarantees with single-server PIR.
Further, an \NAME cluster with 16 \NAME systems achieves 413.0, 544.6, and 127.5 QPS for $\mathtt{Vcall}$, $\mathtt{Comm}$, and $\mathtt{Fsys}$, respectively, with a batch size of 128.
Compared to INSPIRE~\cite{isca-2022-inspire}, which uses in-storage ASICs for HE-based PIR, \NAME delivers 1,243$\times$ higher QPS per system on average by exploiting $\sim$1K$\times$ greater arithmetic throughput enabled by the high bandwidth of DRAM.
Notably, the \NAME cluster achieves a latency of 0.24s for the $\mathtt{Comm}$ workload, which is 150$\times$ faster than INSPIRE’s single-query processing of 36s, despite batching.

\begin{figure}[t]
     \centering
     \includegraphics[width=0.78\columnwidth]{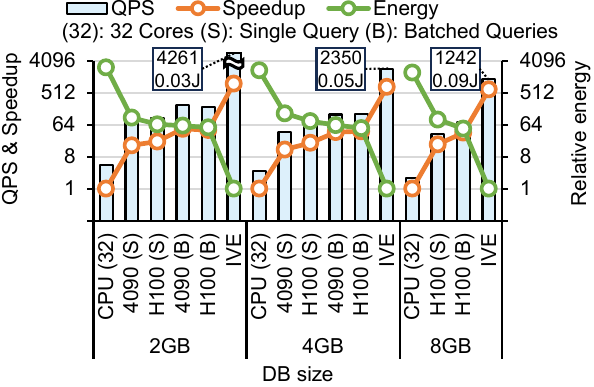}
     \caption{Comparison of PIR throughput and energy consumption (J/query) between \NAME, OnionPIRv2~\cite{iacr-2025-onionpirv2}, and our GPU implementation of OnionPIRv2.}\label{fig:DB_sweep}
\end{figure}

\setlength{\tabcolsep}{2pt}
\begin{table}[t]
\centering
\caption{Number of PIR queries handled in a second (QPS) of \NAME and that of prior PIR hardware acceleration studies.}\label{tab:comparison}
\label{tab:transposed}
\begin{tabular}{c|c|ccccc}
\toprule
 \multicolumn{2}{c|}{}  & \makecell{CIP-PIR\\\cite{usenixsec-2022-gpupir}\textsuperscript{\ddag}} 
 & \makecell{DPF-PIR\\\cite{github-gpudpf}\textsuperscript{*}} 
 & \makecell{INSPIRE\\\cite{isca-2022-inspire}\textsuperscript{\ddag}} 
 & \NAME 
 & \makecell{per \NAME system \\ (vs. INSPIRE) } \\
\midrule
\multicolumn{2}{c|}{Server config} & Multi & Multi & Single & Single & - \\
\multicolumn{2}{c|}{Platform} & GPU & GPU & ASIC & ASIC & - \\
\toprule
 Workload & DB size & \multicolumn{5}{c}{QPS} \\
\midrule
\multirow{3}{*}{\makecell{Synthesized\\DB}} 
& 2GB & - & 956 & - & 4,261 & 4,261 (-) \\
& 4GB & 33.2 & 466 & - & 2,350 & 2,350 (-) \\
& 8GB & 16.0 & 225 & - & 1,242 & 1,242 (-) \\
\midrule
$\mathtt{Vcall}$ & 384GB & - & - & 0.021 & 413.0\textsuperscript{\dag} & 25.8 (1229$\times$)\\
$\mathtt{Comm}$& 288GB & - & - & 0.028 & 544.6\textsuperscript{\dag} & 34.0 (1225$\times$)\\
$\mathtt{Fsys}$ & 1.25TB & - & - & 0.006 & 127.5\textsuperscript{\dag} & {\color{white}0}8.0 (1275$\times$)\\
\bottomrule
\end{tabular}
\begin{itemize}[leftmargin=*]\footnotesize
\item[\ddag] We used the reported values in the paper.
\item[*] We evaluated its open-source implementation~\cite{github-gpudpf} on RTX4090.
\item[\dag] We used an \NAME cluster with 16 \NAME systems and 128 batch size.
\end{itemize}
\end{table}
\setlength{\tabcolsep}{6pt}

\subsection{Sensitivity study}
\label{sec:sensitivity_study}

\noindent \textbf{Algorithmic optimizations:}\
Our algorithmic optimizations significantly improve the overall PIR throughput.
For a 16GB \DB (see Fig.~\ref{fig:sensitivity_b}), applying DFS-based HS reduces the execution time of \expandquery and \restdim by 1.3$\times$ and 1.75$\times$, respectively, leading to an end-to-end latency reduction of 1.2$\times$.
Reduction overlapping (R.O.) achieves an additional 1.06$\times$ improvement over DFS-based HS by accelerating \restdim 1.28$\times$.
Finally, combining our optimizations yields an additional 1.26$\times$ reduction in end-to-end latency, improving \expandquery and \restdim by 1.3$\times$ and 2.23$\times$, respectively.

\noindent\textbf{Query batch size:}\
Regardless of \DB size, batching effectively reduces \DB access bandwidth demands, making \firstdim compute-bound.
Fig.~\ref{fig:sensitivity_c} and \ref{fig:sensitivity_d} present the latency breakdown and throughput for a single \NAME (16GB), an \NAME system (128GB), and an \NAME cluster (1TB with 16 systems) across various batch sizes.
For a 16GB \DB fitting in HBM, throughput increases until a batch size of 64, yielding 1.1$\times$ higher QPS than at 32.
Beyond this, the benefits of batching plateau as \initdim becomes compute-bound and other steps remain non-amortizable, resulting in a linear latency growth and a saturation QPS of 591.
For larger 128GB and 1TB {\DB}s, which require LPDDR offloading, our system requires a larger batch size of 128 to reach saturation in order to compensate for the lower bandwidth of LPDDR, achieving 79.9 and 9.89 QPS per system, respectively.
At saturation, the product of QPS per \NAME and \DB size remains nearly constant, showing scalability.

While batching significantly boosts throughput, it incurs only a modest increase in service latency---3.46$\times$, 1.60$\times$, and 1.62$\times$ for 16GB, 128GB, and 1TB {\DB}s, respectively (Fig.~\ref{fig:sensitivity_c} and \ref{fig:sensitivity_d}).
As each query transfers only a few MBs of client-specific data through PCIe, communication overhead (Comm.) is negligible ($<$8\%), dropping below 1\% for larger {\DB}s.
Throughput gains are substantial at 18.9$\times$ (16GB), 79.9$\times$ (128GB), and 79.1$\times$ (1TB), far outweighing the latency increase and underscoring the effectiveness of batching.

\noindent \textbf{Architectural optimizations:}\
Using special primes reduces both energy consumption and system area.
Compared to the baseline (\textit{Base} in Fig.~\ref{fig:sensitivity_e}), which uses separate computing units for (i)NTT and GEMM (each with the same throughput as a sysNTTU), special primes reduce area and energy by 4\% (\textit{+Sp} in Fig.\ref{fig:sensitivity_e}).
Meanwhile, there is a trade-off in adopting sysNTTU.
As all PIR steps are processed sequentially, the system (\textit{+SysNTTU} in Fig.~\ref{fig:sensitivity_e}) incurs no performance loss and achieves a 7\% area reduction. However, energy consumption increases by $1.1\times$ due to extra circuits required to support both (i)NTT and GEMM within a single unit.

\subsection{Applicability of \NAME to other PIR schemes}
\label{sec:evaluation:other_schemes}

Although HE-based PIR schemes differ in structure, they share core HE operations.
SimplePIR~\cite{usenixsec-2023-simplepir} mainly performs modular GEMMs, while KsPIR~\cite{ccs-2024-kspir} and InsPIRe~\cite{insPIRe}, a concurrent PIR work, rely on automorphism, key-switching, and external-products, which are all efficiently supported by \NAME.
Using 2--4 GB {\DB}s, \NAME achieves 1,904--2,063$\times$ and 3,246--3,347$\times$ higher QPS than the CPU versions of SimplePIR and KsPIR (Table~\ref{tab:other_schemes}).
As most HE-based PIR protocols~\cite{sp-2022-spiral, ccs-2024-respire, popets-2023-frodopir, insPIRe} rely on similar primitives, \NAME can potentially support them with minimal changes.

\setlength{\tabcolsep}{4pt}
\begin{table}[t]
\caption{Performance comparison of CPU implementations and IVE for other single-server PIR schemes.}\label{tab:other_schemes}
\centering
\begin{tabular}{c|cc|cc}
\toprule
\multirow{3}{*}{Scheme} & \multicolumn{4}{c}{QPS}             \\ \cline{2-5} 
& \multicolumn{2}{c}{2GB}    & \multicolumn{2}{c}{4GB}   \\ \cline{2-5} 
                        & CPU & IVE                  & CPU & IVE                 \\
\midrule
SimplePIR~\cite{usenixsec-2023-simplepir}               & 6.2 & 11,766 (1,904$\times$) & 2.9 & 5,883 (2,063$\times$) \\
KsPIR~\cite{ccs-2024-kspir}                   & 0.8 & {\color{white}0}2,555 (3,347$\times$)  & 0.4 & 1,288 (3,246$\times$) \\
\bottomrule
\end{tabular}
\end{table}
\setlength{\tabcolsep}{6pt}

\begin{figure}[t]
     \centering
     \subfloat{\label{fig:vs_ark:overall}}
     \subfloat{\includegraphics[width=0.99\columnwidth]{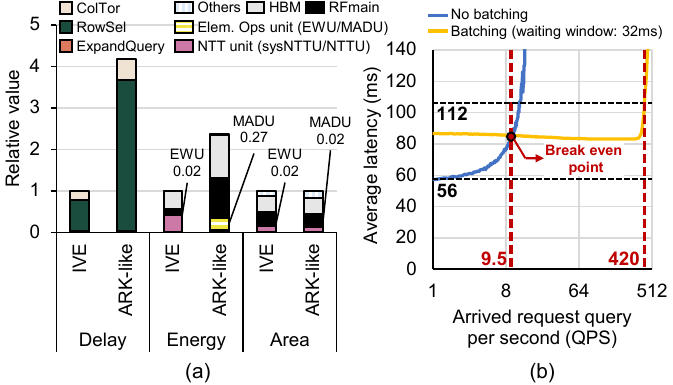}\label{fig:vs_ark:load_latency}}
     \caption{(a) Energy, delay, and area comparison of \NAME and an ARK~\cite{micro-2022-ark}-like system, and (b) load-latency curve under our batch scheduling strategy, both measured with a 16GB \DB.}\label{fig:vs_ark}
\end{figure}

\subsection{Comparison to prior HE accelerator architectures}
\label{sec:evaluation:prior_work}

To evaluate \NAME's hardware efficiency, we compare it against an ARK-like system~\cite{micro-2022-ark}, representative of state-of-the-art HE accelerators~\cite{isca-2022-craterlake, isca-2023-sharp, micro-2021-f1}.
We downscale ARK, which is designed for $N\!=\!2^{16}$, to match our parameters ($N\!=\!2^{12}$).
The ARK-like baseline includes 64 cores, each equipped with an NTTU (matching \NAME's total NTT throughput), two multiply-add units (MADUs)~\cite{micro-2022-ark}, 2MB of scratchpad memory, and other functional units from IVE.
As ARK lacks a dedicated GEMM unit, GEMM operations are mapped to the MADUs. 
For other configurations, to ensure a fair comparison, we applied the same settings, including special-prime usage, on both \NAME and ARK.

\NAME outperforms the ARK-like system in both latency and energy efficiency (Fig.~\ref{fig:vs_ark:overall}).
While the sysNTTUs in \NAME fully exploit the increased arithmetic intensity of batched \firstdim, the ARK-like system suffers from the limited GEMM throughput of the MADUs.
As a result, \NAME achieves 4.2$\times$ higher PIR throughput for a 16GB \DB.
Also, MADU-based GEMM operations require repeated data access to the RF, leading to higher energy consumption.
Moreover, the smaller per-core on-chip memory in ARK limits data reuse, resulting in additional energy consumption due to increased DRAM access.
Consequently, the ARK-like system consumes 2.4$\times$ more energy to retrieve a record from the \DB.
Finally, as the sysNTTUs introduce only a 1.4\% area overhead, the total area of IVE remains comparable to that of the ARK-like system.
Overall, IVE achieves an 9.7$\times$ better energy-delay-area product (EDAP) than the ARK-like system.

\subsection{Batch scheduling system}
\label{sec:eval:batch_schedule}

Even in practical deployment scenarios with random query arrivals and limited concurrency, \NAME with a batch scheduler can accommodate a wide range of load conditions through its high throughput, while restricting the latency overhead to at most 2$\times$ compared to the non-batching case.
We demonstrate this by modeling request arrivals with a Poisson distribution and measuring the PIR latency of a 16GB \DB under different load conditions (see Fig.~\ref{fig:vs_ark:load_latency}).
Our strategy reaches a break-even point at 9.5~QPS, where the average latency of the non-batching case rises sharply as it approaches its throughput limit, which is the reciprocal of the single-query processing rate (17.8 in Fig.~\ref{fig:vs_ark:load_latency}).
Beyond this point, batching keeps the latency within the 2$\times$ bound up to 420~QPS, whereas the latency of the non-batching case continues to increase rapidly, resulting in a 44.2$\times$ throughput advantage.

%% file: related.tex
\section{Related Work}
\label{sec:related_work}

\noindent\textbf{Other HE-based PIR}:\
Numerous studies have proposed PIR protocols based on HE.
SimplePIR~\cite{usenixsec-2023-simplepir} and FrodoPIR~\cite{popets-2023-frodopir} reduce computation by offloading part of the work to an offline phase, but this comes at the cost of client-side storage and preprocessing overhead.
FastPIR~\cite{osdi-2021-fastpir} leverages homomorphic automorphism to reduce computational complexity, and INSPIRE~\cite{isca-2022-inspire} builds on this by optimizing the algorithm to lower query data transfer.
Other studies~\cite{sp-2018-sealpir, security-2021-mulpir, ccs-2021-onionpir, sp-2022-spiral, ccs-2024-respire} focus on reducing communication overhead while balancing computational cost.
Some approaches~\cite{sp-2018-sealpir, security-2021-mulpir} use query compression, accepting additional server-side cost for decompression.
Others~\cite{ccs-2021-onionpir, sp-2022-spiral, ccs-2024-respire} mitigate HE-induced data expansion through external product.

\noindent\textbf{GPU/ASIC acceleration of PIR}:\
Despite growing interest in PIR, only a few studies~\cite{ASPLOS-2024-gpupir, usenixsec-2022-gpupir, isca-2022-inspire} have explored hardware acceleration using GPUs or ASICs, each targeting different PIR protocols.
\cite{ASPLOS-2024-gpupir} focuses on two-server computational PIR based on distributed point functions (DPFs), rooted in multi-party computation, while \cite{usenixsec-2022-gpupir} enhances multi-server information-theoretic PIR.
Most relevant to our work, INSPIRE~\cite{isca-2022-inspire} proposes an in-storage accelerator for HE-based single-server PIR.
These studies highlight the diversity of PIR acceleration strategies, each addressing distinct challenges and optimization opportunities specific to their target protocol.

\noindent\textbf{Alternative privacy solutions:}\
PIR is often compared to oblivious RAM (ORAM), which similarly enables private access to a database.
For example, Compass~\cite{osdi-2025-compass} builds a semantic search system based on ORAM.
While it typically incurs $\mathcal{O}(\log D)$--$\mathcal{O}(\log^2 D)$ overhead per data access~\cite{jacm-2018-pathoram, usenix-2015-ringoram}, its design focuses on concealing access patterns for encrypted private data.
In contrast, PIR works over public or shared data for multiple clients.
While numerous studies~\cite{sp-2018-oblix, ndss-2018-zerotrace, usenix-2023-enigmap, usenix-2025-h2o2ram} use trusted execution environments (TEEs) for enhanced performance, its fundamental risks lie in microarchitecture-based attacks, as continue to be discovered in~\cite{usenix-2018-foreshadow, eurosp-2019-sgxpectre, eurosec-2017-cache-attack, woot-2017-sgx-cache, sp-2020-plundervolt, ccs-2021-smashex}.
Meanwhile, anonymous networks, such as Tor~\cite{usenix-2004-tor} and dummy queries~\cite{sp-2012-ob-pws, 2009-hkpir, tit-2010-query-forgery} can be used to provide privacy; Wally~\cite{arxiv-2024-wally} uses both of them to implement private semantic search.
Although these alternative solutions often achieve asymptotically lower costs than PIR, they mostly rely on additional assumptions or provide privacy guarantees less rigorous than those offered by cryptographic PIR.

%% file: conclusion.tex
\section{Conclusion}
\label{sec:conclusion}

We have proposed \NAME, a hardware accelerator for single-server private information retrieval (PIR) based on homomorphic encryption (HE).
We identified that, after mitigating the memory access burden from database reads via batching, the memory bandwidth demand for client-specific data access emerges as the primary bottleneck, limiting overall throughput.
In response, \NAME employs a large on-chip scratchpad memory and an optimized binary tree search algorithm that maximizes on-chip reuse of the data, thereby achieving high throughput.
We also introduced a versatile functional unit capable of handling major computations in PIR, improving the area efficiency of \NAME.
We proposed a scalable deployment system that combines a scale-up design of \NAME using a heterogeneous memory hierarchy and a distributed scale-out cluster.
As a result, \NAME delivers up to 1,275$\times$ higher throughput compared to prior hardware solutions for HE-based PIR on practical workloads.

%% file: acknowledgment.tex
\section*{Acknowledgment}
\label{sec:acknowledgment}

This research was in part supported by Institute of Information \& communications Technology Planning \& Evaluation (IITP) grant funded by the Korea government (MSIT) [RS-2021-II211343, RS-2024-00469698, RS-2025-02217656, RS-2025-02304125].
The EDA tool was supported by the IC Design Education Center (IDEC), Korea.
This work was conducted while Sangpyo Kim was with Seoul National University (SNU).
Hyesung Ji and Jongmin Kim are with the Interdisciplinary Program in Artificial Intelligence, SNU.
Wonseck Choi and Jaiyoung Park are with the Department of Intelligence and Information, SNU.
Jung Ho Ahn, the corresponding author, is with the Department of Intelligence and Information and the Interdisciplinary Program in Artificial Intelligence, SNU.